\documentclass[12pt]{article}
\usepackage[backend=bibtex, style = chicago-authordate, isbn=false, doi=false, url=false]{biblatex}

\AtEveryBibitem{\clearfield{pages}} % drop pages # in the reference

\usepackage{eurosym} % for euros symbol
\usepackage{amsfonts}
\usepackage{fancyhdr}
\usepackage[usenames,dvipsnames,svgnames,table]{xcolor}
\usepackage[hypertexnames=true]{hyperref} %This makes hyperref ``dumber'', and, hence, more robust! (otherwise sometimes the appendix links don't work).
\usepackage[pdftex]{graphicx}
\usepackage{amsmath, amsthm, amssymb, dsfont, amsfonts}
\usepackage[american]{babel}
\usepackage{color}
\usepackage{morefloats}
\usepackage{tabulary}
\usepackage{tabularx}
\usepackage{booktabs}
\usepackage{fullpage}
\usepackage{setspace}
\usepackage{float}
\usepackage{pdfpages}
\usepackage{lscape}
\usepackage{multirow}
\usepackage{array}
\usepackage{sectsty}
\usepackage{pdflscape}
\usepackage{placeins}
\usepackage[font={large,sc}]{caption}
\usepackage{comment}
\usepackage[margin=1in,headsep=.4in]{geometry}
\usepackage[normalem]{ulem}
\usepackage{tikz}
\usepackage{tikzscale}
\usepackage{xr}
\usepackage[figuresright]{rotating}
\usepackage{subcaption}
\usepackage{caption}
\usepackage{makecell}
\usepackage{graphicx}
\usepackage{hyperref}
\usepackage{pdfpages}
\usepackage{afterpage}
\usepackage{varioref}
\usepackage{ulem}

\definecolor{dark-red}{rgb}{0.4,0.15,0.15}
\definecolor{dark-red}{rgb}{0.55,0.15,0.15}
\definecolor{dark-blue}{rgb}{0.15,0.15,0.4}
\definecolor{medium-blue}{rgb}{0,0,0.5}
\hypersetup{
	colorlinks, linkcolor={dark-red},
	citecolor={dark-red}, urlcolor={dark-red}
}

\labelformat{definit}{ Definition~#1 }

\labelformat{prop}{ Proposition~#1 }

\labelformat{lem}{ Lemma~#1 }
\newtheorem*{proof*}{Proof}

\usepackage{titlesec}
\titlespacing*{\section}{0pt}{1.5ex plus 1ex minus .2ex}{0.8ex plus .2ex}
\titlespacing*{\subsection}{0pt}{1.2ex plus 1ex minus .2ex}{0.8ex plus .2ex}

\usepackage{mathpazo} %Use Palotino fonts
\newcommand*{\FigurePath}{Figures/}

\title{The Fiscal Cost of Public Debt and Government Spending Shocks\footnotetext{\vspace{-0.7cm} \protect\singlespacing\protect \noindent University of California, Berkeley and Sciences Po Paris. v.riblier@berkeley.edu \newline I thank Nicolas Coeurdacier, Yuriy Gorodnichenko, Stéphane Guibaud, Théodore Humann, Emi Nakamura, Xavier Ragot for helpful comments.   }}

\author{Venance Riblier}

\AtEveryBibitem{\clearfield{number}}
\AtEveryBibitem{\clearfield{volume}}

\usepackage{threeparttable}

\begin{document}
 \maketitle

 \begin{abstract}
    This paper investigates how the cost of public debt shapes fiscal policy and its effect on the economy. 
    Using U.S. historical data, I show that when servicing the debt creates a fiscal burden, the government responds to spending shocks by limiting debt issuance. As a result, the initial shock triggers only a limited increase in public spending in the short run, and even leads to spending reversal in the long run. 
    Under these conditions, fiscal policy loses its ability to stimulate economic activity. 
    This outcome arises as the fiscal authority limits its own ability to borrow to ensure public debt sustainability. 
    These findings are robust to several identification and estimation strategies. 


    \vspace{-0.2cm}
 \end{abstract}

\newpage
\FloatBarrier
\section{Introduction}
\label{Introduction}

    Understanding the impact of fiscal policy on economic activity is a central question for policymaking.
    An active strand of research has addressed this issue, with a particular emphasis on quantifying the fiscal multipliers.  
    In these studies, macroeconomists' attention is mostly focused on debt-financed stimulus packages, so as to disentangle the spending and the tax multipliers. 
    Yet, the initial ability of the government to issue debt, as well as the fiscal pressure associated with the service of the debt can vary greatly with the state of the economy. 
    This, in turn can affect both the conduct and the effectiveness of fiscal policy. 
    Indeed, interest payments tighten the government budget constraint, but also reduce the scope for further debt financing, insofar as the fiscal authority is concerned with debt sustainability.
    Therefore, this paper seeks to assess the extent to which the initial cost of servicing the debt influences fiscal policy and its subsequent impact on the economy.

    In contrast to the literature, this paper investigates not only the state-dependence of the fiscal multipliers, but also of fiscal policy itself. 
    Empirical research on fiscal multipliers typically defines fiscal policy as an exogeneous stochastic process for public spending, whose innovations are seen as spending shocks.
    In such models, the fiscal multipliers can be state-dependent, while the fiscal policy itself is not. 
    By contrast, I propose to adopt an alternative view, where fiscal policy is an endogenous variable set by the government, and thus can vary with the initial fiscal conditions. 
    In the underlying model, structural shocks can be thought as shocks to the preference for public spending, rather than shocks to public spending itself. 
    For instance, in the event of a war, it is the greater preference for military equipment, not the actual outlays, that constitutes a spending shock.  
    Considering the path of public spending as endogenous, the question arises as to whether other variables affect the response of the government to the initial shock? 
    In particular, it can be conjectured that among the objectives of the fiscal authority is the sustainability of the debt. 
    Under this hypothesis, the interest paid by the government, which is the main determinant of the trajectory of public debt, also becomes a key factor shaping the response to a spending shock. 

    To test this hypothesis, I estimate the state-dependent multipliers and impulse responses of public spending and output to identified spending shocks. 
    The analysis exploits the large variations of the U.S. historical data.
    To construct a state variable that captures changes in the fiscal burden associated with debt service, I collect security level data from the Monthly Statement of Public Debt. 
    I measure the cost of the debt as the interest charge over every outstanding federal security, expressed in terms of GDP, and compare the resulting series to its historical trend to define episodes of large debt cost.

    The main findings of this paper is that fiscal policy is strongly state-dependent with the initial burden of servicing the debt. 
    When the cost of the debt is large, the response of government spending to a shock is short-lived and quickly reverts to zero. 
    At long horizon, a positive shock even results into a contraction of public spending.
    By contrast, when the cost of servicing the debt is limited, the response of government spending is much more important and persistent, and reverts to zero only at long horizon.
    This pattern can be explained by the dynamics of public debt following the initial shock. 
    When it hits the economy during episodes of large debt cost, the fiscal authority limits debt issuance.
    As the response of tax revenues does noy vary greatly over states, the only margin of adjustment left in the government budget constraint is public spending.

    This asymmetry in fiscal policy is determinant for the response of output.
    To illustrate this idea, one can reformulate the standard cumulative multipliers, defined as the ratio of the cumulative change in output to the cumulative change in spending (\cite{mountfordWhatAreEffects2009}), as
    \begin{equation}
        \sum_{i=0}^h \Delta y_{t+i} = m_{h}  \sum_{i=0}^h \Delta g_{t+i}
    \end{equation}
    The total change in output resulting from a spending shock can be decomposed into an intensive margin effect: the multipliers $m_h$, and an extensive margin effect: the total change in government spending. 
    A key result of this paper is that the response of output to a spending shock is also contingent upon the initial cost of debt.
    This state-dependence primarily arises from the shape of fiscal policy.     
    The multipliers play a lesser role in the asymmetry of the output response. 
    This finding contributes to the literature, which usually focus only on the multipliers, by showing that the response of output is also strongly influenced by the dynamic response of public spending. 
    It is therefore critical to consider fiscal policy as being endogenous and to analyse its response to shocks to the preference for public spending.

    This paper expands the literature on the financing of fiscal policy in several ways. 
    \cite{perottiFiscalPolicyGood1999}, \cite{ilzetzkiHowBigSmall2013}, \cite{nickelFiscalStimulusTimes2014} and \cite{huidromWhyFiscalMultipliers2020} document a negative relation between large debt-to-GDP ratios and fiscal multipliers.
    The only exception is \cite{auerbachFiscalStimulusFiscal2017} who find no such relation.
    A first contribution of this paper is to highlight the central role of the cost rather than the quantity of public debt in the analysis of fiscal policy. 
    Secondly, I find that higher debt-to-GDP ratios are associated with smaller multipliers only on impact of the shock, with little support for a significant relation at longer horizon. 
    I argue that evidence for such a relation at long horizon might stem from an identification issue.
    This paper is also closely related to the analysis of \cite{corsettiFiscalStimulusSpending2012}, who show that on the period 1983-2007 spending shocks in the U.S. are followed by large public debt build-ups that eventually result into spending reversals. 
    With an extended time sample, I show that such a mechanism arises only when the fiscal pressure of servicing the debt is large. 

    This paper also informs the discussion of public debt sustainability. 
    I show that the asymmetric shape of fiscal policy is the decision of the fiscal authority itself, not the outcome of market or monetary authority constraints.
    When the cost of servicing the debt is large, the Congress reduces the debt ceiling, capping the ability to finance further spending through debt.  
    This findings echoes the seminal paper of \cite{bohnBehaviorUSPublic1998} who show that over the 20th century the U.S. public debt has been sustainable as the fiscal authority responded to rising debt by cutting deficits.
    This paper highlights the shadow cost behind \cite{bohnBehaviorUSPublic1998} result. 
    To achieve public debt sustainability, the government has to prevent large build-ups when the service of the debt is already burdening.
    This requires to sacrifice fiscal policy during such episodes, which is harmful to output.

    Eventually, this paper presents methodological contribution to the literature on fiscal policy. 
    Traditionally, the estimation of impulse responses is carried out through VAR or local projections, with the latter being more suitable for dealing with state-dependence or cumulative multipliers (see \cite{rameyGovernmentSpendingMultipliers2018}).
    Yet, as \cite{liLocalProjectionsVs2022} have shown, neither strategy is the optimal solution to the bias-variance trade-off. 
    As a result, I conduct the estimation using the smooth local projections method introduced by \cite{barnichonImpulseResponseEstimation2019}. 
    Smooth local projections expand the traditional local projections à la \cite{jordaEstimationInferenceImpulse2005}  by shrinking the impulse responses toward to polynomial functions of the time horizon. 
    I show that such a shrinkage estimator achieves a substantial reduction in the noise of the impulse responses, while introducing only a minor bias. 

    Secondly, this paper introduces a new series of identified spending shocks. 
    I apply the narrative sign restrictions method proposed by \cite{antolin-diazNarrativeSignRestrictions2018} to the U.S. data on public spending.
    This approach is less restrictive than the traditional timing restriction proposed by \cite{blanchardPerottiQJE2002} and yields a narrower set of identified shocks compared to the standard sign restrictions used by \cite{mountfordWhatAreEffects2009}.
    I also employ the usual narrative news shocks of \cite{rameyIdentifyingGovernmentSpending2011} and the timing restriction approach of \cite{blanchardPerottiQJE2002} and show that the results are robust to the three identification strategies. 
    
    The rest of the paper is organized as follows. Section \ref{section Public Debt and Spending Shocks in the U.S.} describes the data, the construction of the state variable and the identification of the spending shocks. In section \ref{section Empirical Strategy}, I introduce the empirical strategy. Section \ref{section Fiscal Policy and the Cost of Public Debt} presents the main results of the paper. In section \ref{section Sensitivity Analysis}, I discuss the robustness of these findings. Section  \ref{section Conclusions} concludes.

\FloatBarrier
\section{Public Debt and Spending Shocks in the U.S.}
\label{section Public Debt and Spending Shocks in the U.S.}

    \FloatBarrier
    \subsection{Data and Definition of State Variable}
    \label{subsection Data Description and state variable}

        To investigate whether the effect of government spending shocks depends on the fiscal cost of public debt, I use historical data for the U.S. 
        The data are observed at the quarterly frequency and cover the period 1889Q1-2015Q2.
        Most series are from \cite{rameyGovernmentSpendingMultipliers2018}, such as the NIPA variables including GDP, government spending and revenues, but also the GDP deflator and the estimates of potential GDP.
    
        I expand the data set by constructing a time series for the interest bearing federal debt and a measure of the cost associated with this debt. This metrics is defined as the sum of the interest charge on every federal security outstanding, expressed in terms of GDP. That is:
        \begin{equation}
            \text{fiscal cost}_t = \frac{\sum_{i} b_{i,t} r_{i,t}}{GDP_t}
        \end{equation}
       
\begin{figure}[h!]
            \caption{The Fiscal Cost of the U.S. Public Debt}
            \vspace{-0.2cm}
            \begin{center}
              \includegraphics{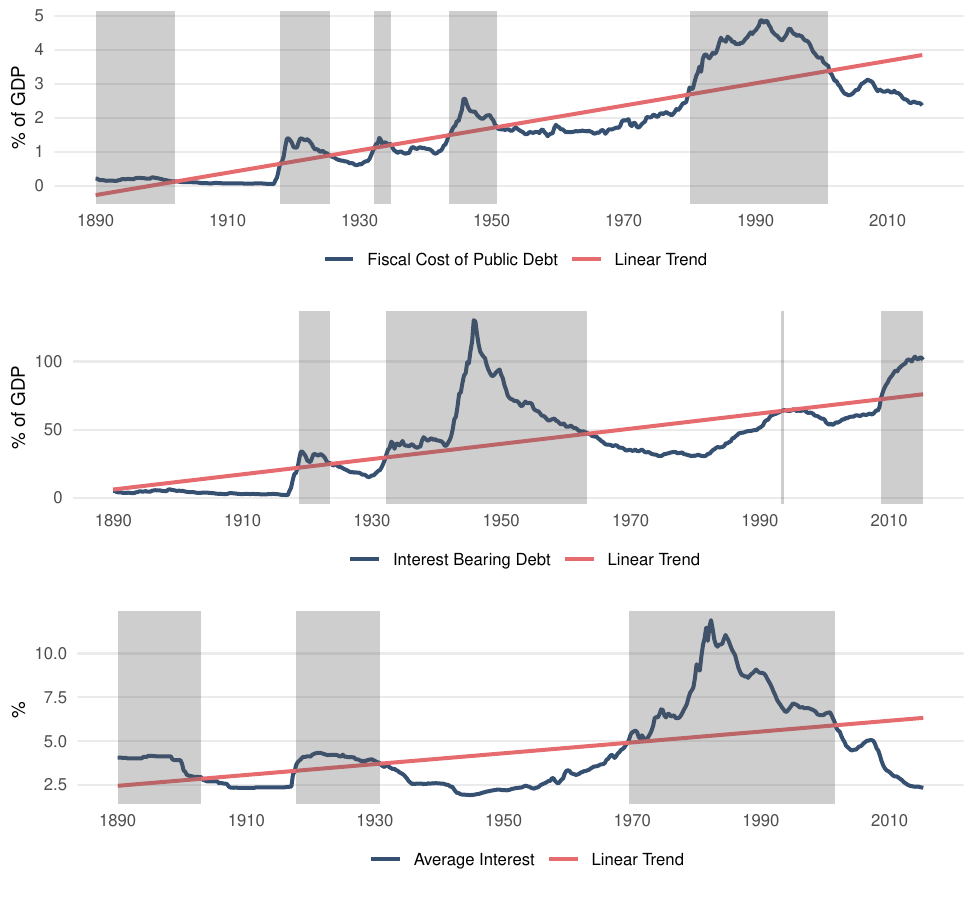} 
              \label{fig ts_debt_cost_state}
            \end{center}
            \vspace{-1.3cm}
            \begin{small}     
              \textit{Note:} This figure shows the fiscal cost of public debt and its components. The fiscal cost of public debt is measured as the sum of the interest charge over outstanding securities. The average interest rate is obtained by dividing the fiscal cost by the quantity of interest bearing debt. 
              Shaded areas define periods where a series is larger than its linear trend.
            \end{small}
          \end{figure}

        Where $b_{i,t}$ is the amount of security $i$ outstanding in $t$ and $r_{i,t}$ is the annual interest rate that the Treasury pays on the security. To construct this metrics, I collect security level data on $b_{i,t}$ and $r_{i,t}$ from the Monthly Statement of Public Debt. 
        The upper panel of figure \ref{fig ts_debt_cost_state} shows the resulting time series. 
        It is clear from this figure that historical data offers substantial variations in the fiscal cost of public debt\footnote{One could wonder why focusing on the U.S. over a panel approach?
        I argue that a cross-section dimension is not likely to improve the identification of the different states. With a panel of advanced economies that excludes default risk, the interest rates on the different countries are highly correlated so cross-section variations in the cost of debt boil down to variations in the quantity of debt. 
        On the other hand, including more countries would make states of costly debt hard to interpret as they capture not only the fiscal burden faced by the governments, but also the crisis faced by defaulting or high spread countries. 
}, which is key to isolate different states of the economy and further identify how they interact with spending shocks.

        To define the states where the fiscal burden of public debt is large, I compare the cost of debt to its historical average, controlling for the time trend. 
        This yields five episodes of costly debt, which can be seen on the shaded areas in the upper panel of figure \ref{fig ts_debt_cost_state}. 
        Arguably, these states capture periods when servicing public debt put a strain on public finance, and are not confounded with other potential state variables.
        In particular, there is no systematic link between between the total cost of public debt and one of its components (shown in the two lower panels of figure \ref{fig ts_debt_cost_state}).  
        Two episodes (1889Q1:1902Q1 and 1980Q1:2001Q1) are mostly driven by the average interest rate, two others by the debt-to-GDP ratio (1932Q2:1934Q4) and (1943Q3:1950Q4) and the last one (1918Q1:1925Q3) appears to be equally explained by the large quantity of debt and average interest rate.
        This is an appealing feature of historical data that allows to address specifically the state-dependence in the fiscal cost of public debt, apart from the potential effect of the quantity of debt. 

\begin{figure}[h!]
          \caption{States of Costly Debt: Smoothed Probability}
          \vspace{-0.2cm}
          \begin{center}
            \includegraphics{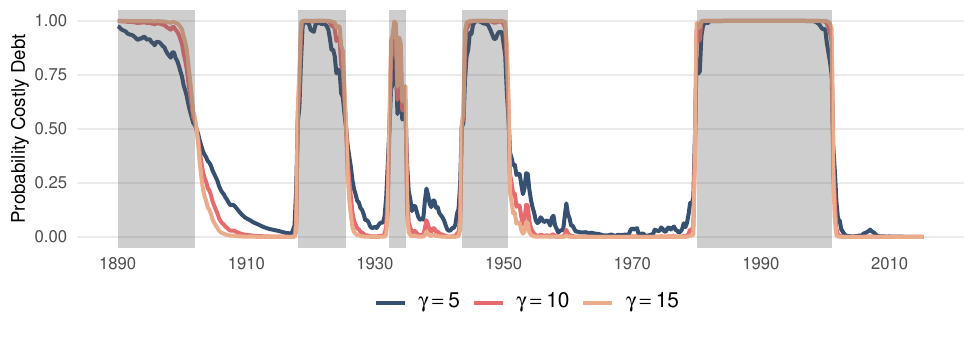} \label{fig ts_smooth_states}
          \end{center}
          \vspace{-1.4cm}
          \begin{small}
            \textit{Note:} Shaded areas show states where the cost of public debt is larger than its linear trend. Solid lines show the logit probability of the economy being in state of costly debt for different speed of transition $\gamma$. 
          \end{small}
        \end{figure}

        It is clear from figure \ref{fig ts_debt_cost_state} that the cost of debt is a slow moving variable. There is therefore no reason to believe that the economy transitions toward a state to another in only one quarter. To address this concern, I follow  \cite{tenreyroPushingStringUS2016} and compute the probability of being in the fiscal burden state using the logit function:
        \begin{equation}
        \label{eq logit}
        F(x_t) = \frac{1}{1 + e^{-\gamma \frac{x_t}{\sigma (x_t)}}}
        \end{equation}
        Where $x_t$ is the detrended cost of debt, $\sigma(x_t)$ its standard deviation and $\gamma$ a parameter governing the speed of the of the transition. The smaller $\gamma$ the smoother the transition between two states. 
        Figure \ref{fig ts_smooth_states} shows the probability of being in a state of costly debt for different values of $\gamma$. 
        It illustrates that the speed of transition changes between different episodes of costly debt. 
        For instance, the transition in 1902Q1 is much smoother than in 1980Q1. 
        Too large value of $\gamma$ would eliminate this information, but too small values would overstate the smoothness of the transition. As a result, I use an intermediate value of $\gamma = 10$ as baseline, and show the robustness of the results when using different values as well as a dummy state variable in appendix \ref{appendix subsection smoothness of transition}.

        There are several alternatives for defining the state variable. A first option is to compare the cost of debt to a flexible time trend, such as that obtained with an HP filter or with a moving average over narrow time window. Although this approach gives more variations in the state variable, it can lead to misleading interpretations.
        For instance, a slight decrease from an initially high cost of debt might be identified as a change in the state of the economy, while the fiscal pressure from public debt remains overwhelming. 
        I assess the robustness of the results to this approach in appendix \ref{appendix subsection Increase state frequency}. 
        Alternatively, one can adopt a less parametric approach, by using directly the cost of debt in level as a continuous state variable. 
        I apply and discuss this method in section \ref{subsection continuous state variable}

    \FloatBarrier
    \subsection{Identifying Government Spending Shocks}
    \label{subsection Identifying Government Spending Shocks}

        The literature has reached a consensus on two approaches for identifying exogenous changes in government spending. 
        These are using a timing restriction in a structural VAR as \cite{blanchardPerottiQJE2002} and the narrative approach proposed by \cite{rameyIdentifyingGovernmentSpending2011}.

        The popularity of \cite{rameyIdentifyingGovernmentSpending2011} news shocks is primarily due to the credibility of its exogeneity. 
        The instrument is constructed using periodical evidence that captures the exact timing of the news on government spending and thus ensure that the shocks are unanticipated.
        To rule out reverse causality, the author focuses on changes in spending caused by foreign political events, like wars. This strong exogeneity comes at a cost of limited relevance, as noted by \cite{rameyGovernmentSpendingMultipliers2018}. 

        By contrast, the timing restriction shocks estimated following \cite{blanchardPerottiQJE2002} approach are known to be more relevant instrument to estimate fiscal multipliers. This identification assumes that government spending does not contemporaneously respond to innovations in output, which can be questioned, even at the quarterly frequency.
        For instance, in response to the pandemics, the CARES act has been signed on March $27^{th}$, the same month that economic activity contracted (see \cite{chettyEconomicImpactsCOVID192020} for high frequency evidence). This provides a compelling case where the fiscal policy can adjust to output shocks in less than a quarter.

        To overcome this limitation, I propose to exploit a novel identification strategy introduced by \cite{antolin-diazNarrativeSignRestrictions2018}. The core of this method is to augment the traditional sign restrictions with narrative information. Standard sign restrictions achieve set identification of a structural shock by constraining the sign of its effect for a certain period, typically one year.
        \cite{antolin-diazNarrativeSignRestrictions2018} show that this identification can be sharpened by imposing further restrictions, directly on the sign of the structural shocks around a handful of historical episodes. 
        In addition, it is possible to impose restrictions on the historical variance decomposition of the structural shocks around these episodes.

        I apply this methodology to identify government spending shocks, which is a novelty in the literature on fiscal policy. Specifically, I use the two quarters when the U.S. enter the World Wars as historical events to discipline the structural shocks. 
        As suggested by the instrument of \cite{rameyIdentifyingGovernmentSpending2011}, these two episodes are undoubtedly instances of significant spending shocks. 

        The shocks are identified by combining three set of restrictions. First, I impose a standard sign restriction that requires any spending shock to have a positive effect on government spending for at least one year. 
        The second restriction applies directly on the structural shocks that hit in 1917Q2 and 1941Q4, requiring them to be positive. Eventually, I constrain the magnitude of these two shocks, by imposing that the absolute value of their contribution to the unexpected change in government spending is larger than the contribution of the sum of other structural shocks.  

        To recover the shocks, I follow \cite{antolin-diazNarrativeSignRestrictions2018} and estimate a Bayesian SVAR, including real GDP, government spending, tax and debt, with four lags. The second step is to draw from the posterior distribution of structural parameters and check for restrictions. Out of 50,000 initial draws, 12,023 satisfy the standard sign restrictions, and 1,847 are consistent with both the standard and the narrative sign restrictions. 
   
        This methodology achieves a much sharper identification, as it narrows down the set of structural parameters consistent with the data. Yet, it still results into set-identification. Thus, I summarize the set of shocks by its median, resulting in one time series of shocks. 
        In the following, I employ this series as well as the standard narrative and timing restriction approaches as identified shocks.  
        To ease the comparisons of the results between the different shocks, I normalize them by their standard deviation.
        For timing and narrative sign restriction shocks, this is equivalent to scale them as having a contemporaneous impact on public spending equal to 1\% of GDP in the linear case.

\FloatBarrier
\section{Empirical Strategy}
\label{section Empirical Strategy}

    \FloatBarrier
    \subsection{Empirical Model}
    \label{subsection Empirical Model}

        To estimate the impulse responses to a spending shock, I use local projections (\cite{jordaEstimationInferenceImpulse2005}). This method is popular for the estimation of fiscal multipliers, especially when allowing for state-dependence, such as in \cite{rameyGovernmentSpendingMultipliers2018}, \cite{bernardiniPrivateDebtOverhang2018}, \cite{barnichonUnderstandingSizeGovernment2022} or \cite{bronerFiscalMultipliersForeign2022}. 
        The local projections method estimates the IRF by projecting the variable of interest $z_{t+h}$ at horizon $h$ onto the information set of period $t$, and thus does not require to make assumptions on how the economy transitions from a state to another between $t$ and $t+h$. 
        By contrast, estimating the response at horizon $h$ with a VAR means compounding $h$ times the one-period ahead forecast. In the nonlinear case, this requires tracking the evolution of the state variable, and thus taking a stand on how it responds to both endogenous and exogenous shocks. Hence, local projections offer a more flexible framework for estimating nonlinear IRF. The empirical model can be written as:  
        \begin{equation}
            \label{eq state dependent lp}
            z_{t+h} = I_{t-1} \left(\alpha_h^A + \beta_h^A shock_t + \phi_h^A(L) X_{t-1}\right) + (1-I_{t-1}) \left(\alpha_h^B + \beta_h^B shock_t + \phi_h^B(L) X_{t-1}\right) + \varepsilon_{t+h}
        \end{equation}
        Where $z_{t+h}$ is the variable of interest at horizon $h = 0, ..., H$ and $I_{t-1}$ a  variable capturing the state of the economy prior to the shock. The impulse response functions in states A and B are given by the collection of coefficients $\beta_{h}^A$ and $\beta_{h}^B$.
        The model also includes a set of lagged controls $ \phi_h^i(L) X_{t-1}$ where $\phi_h^i(L)$ is the lag operator in state $i \in \{A, B\}$ and $X_{t-1}$ a vector comprising real output, government spending, tax and debt. The number of control lags is set to 4. 
        Standard errors are computed following \cite{neweySimplePositiveSemiDefinite1987} to correct for their potential autocorrelation, which is a common concern in local projections.
        To address stationarity, I apply the \cite{gordonEndGreatDepression2010} transformation to the four real variables used in the model, that is to divide them by an estimate of potential GDP.

        I use the model \eqref{eq state dependent lp} to evaluate the IRF of output and government spending to an initial spending shock. 
        In addition, to gauge the size of the fiscal multipliers, I follow \cite{rameyGovernmentSpendingMultipliers2018} and use a local projection IV (LP-IV) model: 
        \begin{equation}
            \label{eq state dependent multiplier}
            \begin{array}{cl}
                \sum_{h=0}^H y_{t+h} = & I_{t-1} \left(\alpha_h^A + m_h^A \sum_{h=0}^H g_{t+h} +  \phi_h^A(L) X_{t-1}\right)  +  \\
                & (1-I_{t-1}) \left(\alpha_h^B + m_h^B \sum_{h=0}^H g_{t+h} + \phi_h^B(L) X_{t-1}\right) + \varepsilon_{t+h}        
            \end{array}
        \end{equation}
   
        Where $y$ specifically refers to output and $g$ to government spending. 
        In this model, the multiplier is derived as the marginal effect of cumulative government spending on cumulative output, with the latter being instrumented by the identified spending shock. 
        The intuition for this approach is as follows. The fitted values of the first stage capture the total amount of government spending between $t$ and $t+h$ that can be explained by the initial shock in $t$.
        Using these fitted values in the second stage, we achieve the definition of the spending multiplier. That is, for all horizon $h$, $m_h^i$ captures the marginal effect of one additional dollar of government spending on output in state $i$, provided that this additional dollar is the outcome of an exogenous shock.
        
        The LP-IV model allows for several instruments. As shown in appendix \ref{appendix section Instrument Relevance for Multipliers}, the \cite{rameyIdentifyingGovernmentSpending2011} news shocks alone are not relevant enough, but they provide valuable information at long horizon when combined with shocks identified in SVAR. Thus, I use two set of instruments to estimate the multipliers. The first one combines news and timing shocks, while the second combines news and narrative sign shocks.
        By contrast, when estimating the IRF of output and spending, only one series of shocks can be used, so the model \eqref{eq state dependent lp} is estimated separately with the three different identified shocks.

    \FloatBarrier
    \subsection{Controlling for the Bias-Variance trade-off}
    \label{section slp}

        As shown by \cite{plagborg-mollerLocalProjectionsVARs2021}, in population, local projections and VARs yield the same estimates. 
        Yet, the two methodologies can be seen as the polar solutions to the bias-variance trade-off. 
        In VARs, the initial error is compounded $h$ times, resulting in a potentially large bias. By contrast, local projections does not impose such structure on the $h$ period ahead forecast, leading to more noisy estimates. 
        \cite{liLocalProjectionsVs2022} show that neither of these estimation strategies is optimal to address the bias-variance trade-off, and suggest to use instead shrinkage estimation. 

        Therefore, I estimate \eqref{eq state dependent lp} and \eqref{eq state dependent multiplier} using smooth local projections. Introduced by \cite{barnichonImpulseResponseEstimation2019},  this method allows to control for the bias-variance trade-off when estimating the impulse response function. 
        The idea behind smooth local projections is to first approximate the IRF coefficients as a linear combination of simple basis functions and then shrink them toward a polynomial using penalized least squares. 
        Imposing a large degree of shrinkage cuts the variance but increases the bias of the estimates as they get closer to a polynomial function of the time horizon. Conversely, with small shrinkage degree, smooth local projections converge to local projections estimates. 

        I detail the estimation procedure in the linear case to keep the presentation clear. Assuming the model to estimate is:

        \begin{equation}
            \label{eq local projection linear}
            z_{t+h} =\alpha_h + \beta_h shock_t + \phi_h(L) X_{t-1} + \varepsilon_{t+h} 
        \end{equation}
        Where, as in section \ref{subsection Empirical Model},  the collection of $\beta_h$ for all $h = 0,...,H$ is the impulse response function of $z$ to the identified shock. The first step of smooth local projections is to approximate $\beta_h$ by a linear combination of simple basis functions:
        \begin{equation}
            \label{eq approx beta_h}
            \beta_h \approx \sum_{k = 1}^K b_k B_k(h)
        \end{equation}
        Where $B_k(h)$ are B-splines basis functions. These basis achieve a smooth approximation of $\beta_h$ and yield well-behaved penalized regressions (see \cite{eilersFlexibleSmoothingBsplines1996}). I follow \cite{barnichonImpulseResponseEstimation2019} and use cubic B-splines basis defined on an equidistant grid with bounds $- 2$ and $H - 1$ (which pins down $K = H + 2$).
        The other coefficients can be approximated the same way, leading to the following model:
        \begin{equation}
            \label{eq lp approximation b spline}
            z_{t+h} \approx \sum_{k=1}^{K} a_k B_k(h) + \sum_{k=1}^{K} b_k B_k(h)  shock_t + \sum_{j=1}^{p} \sum_{k=1}^{K} c_{j,k} B_k(h) x_{j,t} + \varepsilon_{t+h}  
        \end{equation}
        We can stack this linear model in matrix form and write it as:
        \begin{equation}
            \label{eq stacked approx lp}
            \mathcal{Z} = \mathcal{X} \theta + \mathcal{U}
        \end{equation}
        Where $\theta$ is the vector stacking the coefficients $a_{k}$, $b_{k}$ and $c_{j,k}$. It is estimated using penalized least squares:
        \begin{equation}
            \label{eq penalized ls}
            \hat{\theta} = \arg \min_{\theta} (\mathcal{Z} - \mathcal{X}\theta)' (\mathcal{Z} - \mathcal{X}\theta) + \mu \theta' \mathbf{P} \theta
        \end{equation}
        Where $\mathbf{P}$ is the penalty matrix and $\mu$ is the shrinkage degree. 
        To choose the penalty matrix, \cite{barnichonImpulseResponseEstimation2019} exploits an appealing property of B-splines basis, that we can express the derivative of order $r$ of $\sum_{k = 1}^K b_k B_k(h)$ in a simple matrix form $b'D_r' D_r b$, where $D_r$ is the matrix $r^{th}$ difference operator. 
        Assuming that $b$ is ordered first in the vector $\theta$, we can set $\mathbf{P}$ equal to  $D_r' D_r$ for the $(1 : K) \times (1 : K)$ first entries and to 0 elsewhere.
        With this penalty matrix, the estimation shrinks the $r^{th}$ derivative of $\beta_h$ to 0, that is, shrinks $\beta_h$ toward a polynomial of order $r-1$. In the following I use $r=3$, meaning that I shrink the impulse response function toward a piecewise quadratic function of the horizon $h$. As shown in appendix \ref{appendix subsection Degree of Shrinkage} the results are not affected when using larger values for $r$. Eventually, I follow \cite{barnichonImpulseResponseEstimation2019} and use k-fold cross-validation (\cite{racineFeasibleCrossValidatoryModel1997}) to select the optimal value for $\mu$.

\begin{figure}[h!]
          \caption{IRF with Standard and Smooth Local Projections}
          \begin{center}
            \includegraphics{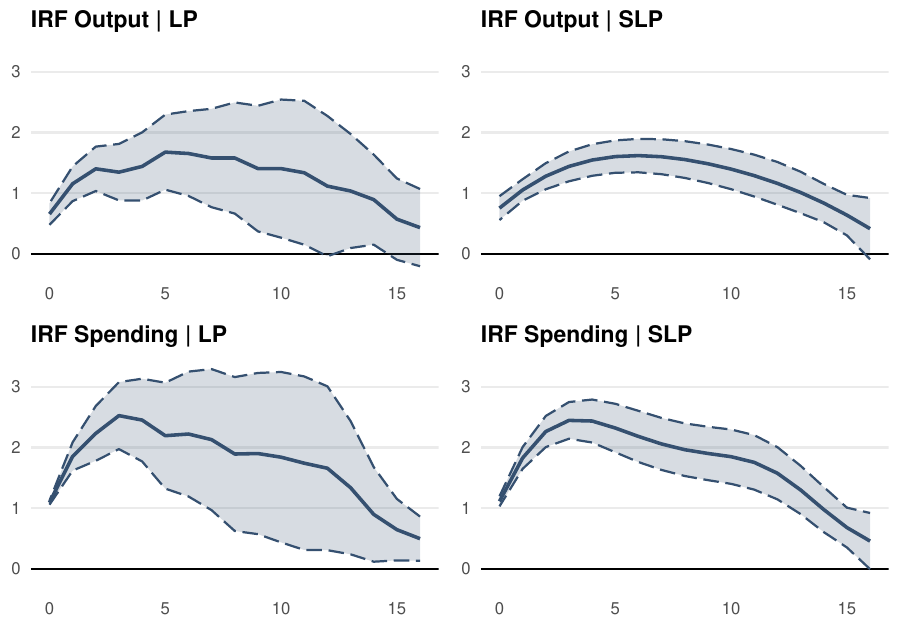} \label{fig linear_lp_slp}
          \end{center}
          \vspace{-0.6cm}
          \begin{small}
            \textit{Note:}  This figure shows the impulse response functions (IRF) to a government spending shock identified with a timing restriction. The two upper panels show the IRF of output, the two lower panels show the IRF of government spending. The IRF on the left panels are estimated using standard local projections (LP). The IRF on the right are estimated using smooth local projections (SLP). The confidence intervals are defined at a 10\% risk level.
          \end{small}
        \end{figure}

        To illustrate the interest of this approach, I compare the IRF of both output and spending, using standard and smooth local projections. Figure \ref{fig linear_lp_slp} shows the results. 
        The gain in variance appears clearly. The size of the confidence interval is roughly divided by two when using SLP instead of LP. Importantly, this gain comes at a small cost of bias. On average, the percentage difference between SLP and LP estimates is 0.56\% for output and 0.16\% for spending. This quantity is negligible compared to the reduction in standard errors when using SLP over LP, which is equal to 52.4\% for output and 38.1\% for spending.

        Therefore, I employ smooth local projections to estimate the model \eqref{eq state dependent lp} for the IRF of output and spending and to estimate the model \eqref{eq state dependent multiplier} for the multipliers.

\FloatBarrier
\section{Fiscal Policy and the Cost of Public Debt}
\label{section Fiscal Policy and the Cost of Public Debt}

    \FloatBarrier
    \subsection{Baseline Results}
    \label{subsection Baseline Results}

\begin{figure}[h!]
            \caption{State-Dependent Responses: Costly Public Debt }
            \begin{center}
                \includegraphics{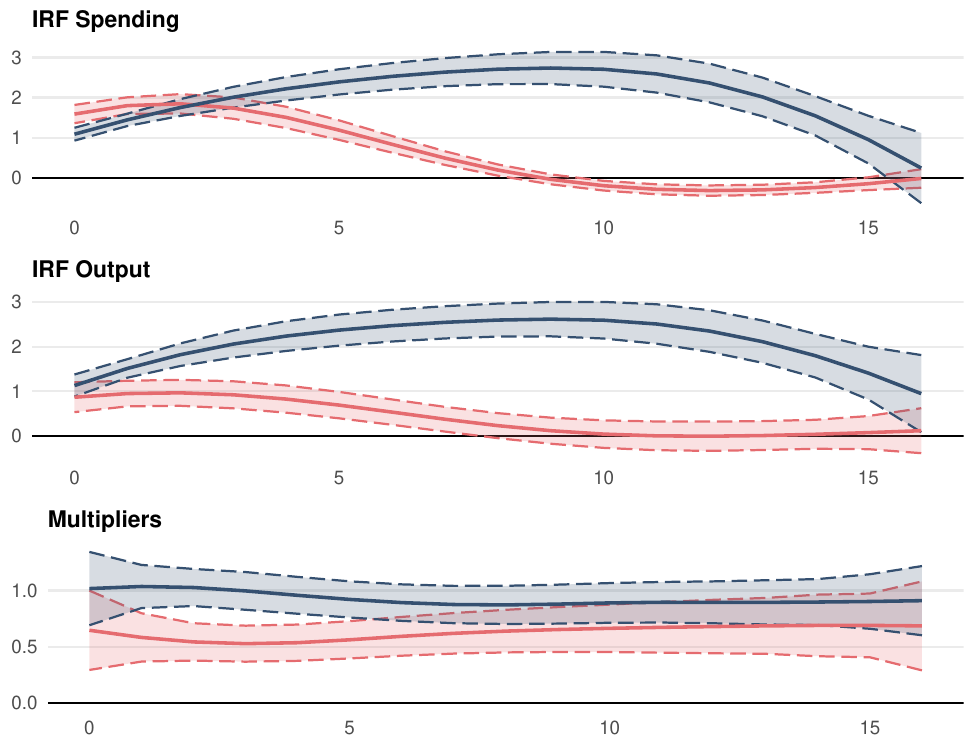} \label{fig baseline_irf}
            \end{center}
            \vspace{-1cm}
            \begin{small}
                \textit{Note:}  This figure shows the state-dependent impulse response functions (IRF) to a government spending shock identified with a timing restriction. 
                The IRF in red (blue) corresponds to initially large (small) debt cost.
                The two upper panels show the IRF of public spending and output, measured in \% of potential GDP. 
                The last panel shows the cumulative multiplier, measured in dollars.
                The IRF are estimated using smooth local projections.
                The forecast horizon is in quarters. 
                The confidence intervals are defined at a 10\% risk level.  
            \end{small}
        \end{figure}

        In this section, I present the main results of the paper.
        Figure \ref{fig baseline_irf} shows the impulse response functions (IRF) of spending and output as well as the cumulative multipliers, in states of large and small debt cost (in red and blue respectively). 
        The impulse responses of output and spending are obtained by estimating the model \eqref{eq state dependent lp} with smooth local projections, using timing restriction shocks. 
        For the multipliers, I estimate the model \eqref{eq state dependent multiplier} with smooth local projections as well, but combine timing restriction shocks with \cite{rameyIdentifyingGovernmentSpending2011} news shocks, as they provide relevant information over the long run. 
        Table \ref{table main_table_baseline} shows the difference in estimates between times of large and small debt service, for the three identification schemes presented in section \ref{subsection Identifying Government Spending Shocks}. The p-values are constructed using HAC standard errors to correct for the serial correlation of residuals.

        The trajectory of government spending varies greatly depending on the state of the economy.
        During episodes of costly debt, the response of spending is larger on impact but much less persistent.
        The positive effect on spending starts to erode after only one year and reverts to zero after two years.
        At long horizon, the initial shock even leads to a reduction of spending. 
        By contrast, when the cost of debt is small, the initial shock has a persistent effect on government spending, which reverts only after four years.
        The gap between the responses is not only statistically significant, but quantitatively important: after two years the difference in point estimates is 2.51 percentage points for a shock identified with timing restriction.

        The IRF of output displays the same state-dependence. 
        When a spending shock hits during a costly debt episode, the response of output is smaller than one and short-lived, it reverts to zero after two years. 
        Conversely, when the cost of debt is small, the initial shock has a large and long-lasting effect on output. 
        It peaks above 2.5\% after two years and is still 1\% after four years. 
        The cumulative multipliers only account for a small share of the difference in output IRF. 
        During the first year after the shock, the multiplier is twice smaller in state of costly debt. This reduces the response of output in the short run, despite the large response of spending in this state. 
        At longer horizon, the difference in multipliers vanishes.
        Then, all the state-dependence in output response is driven by the trajectory of government spending. 

        These results suggest that the IRF of output is strongly state-dependent in the fiscal cost of public debt.
        Most of the difference between states is explained by the short-lived and small response of spending that occurs when public debt induces large fiscal pressure. 
        Smaller multipliers contributes to reducing the response of output only in the short-run. 

            \begin{table}[h!]
                \begin{center}
                \caption{Difference in Estimates: Costly Public Debt}
                \vspace{0.2cm} 
                    \begin{tabular}{@{\extracolsep{10pt}}lccccc}
\toprule
\bottomrule
\\[-1.4ex]
 & $h = 0$ & $h = 4$ & $h = 8$ & $h=12$ & $h=16$ \\
\midrule
& \multicolumn{5}{c}{Timing Res. Shocks} \\ \cline{2-6} \\[-1.8ex]
IRF Spending: & & & & \\
\quad Small Debt Cost & 1.09$^{***}$ & 2.22$^{***}$ & 2.71$^{***}$ & 2.36$^{***}$ & 0.24 \\
\quad Diff. Costly Debt & 0.50$^{***}$ & -0.71$^{***}$ & -2.51$^{***}$ & -2.68$^{***}$ & -0.26 \\
IRF Output:  & & & & \\
\quad Small Debt Cost & 1.13$^{***}$ & 2.24$^{***}$ & 2.60$^{***}$ & 2.35$^{***}$ & 0.95$^{*}$ \\
\quad Diff. Costly Debt & -0.26 & -1.41$^{***}$ & -2.36$^{***}$ & -2.35$^{***}$ & -0.83 \\
\\[-1.4ex]
& \multicolumn{5}{c}{Narrative Sign Res. Shocks} \\ \cline{2-6} \\[-1.8ex]
IRF Spending: & & & & \\
\quad Small Debt Cost & 1.00$^{***}$ & 2.00$^{***}$ & 2.49$^{***}$ & 2.21$^{***}$ & 0.09 \\
\quad Diff. Costly Debt & 0.47$^{***}$ & -0.55$^{**}$ & -2.31$^{***}$ & -2.51$^{***}$ & -0.05 \\
IRF Output:  & & & & \\
\quad Small Debt Cost & 1.40$^{***}$ & 2.54$^{***}$ & 2.69$^{***}$ & 2.40$^{***}$ & 1.10$^{**}$ \\
\quad Diff. Costly Debt & -0.18 & -1.44$^{***}$ & -2.28$^{***}$ & -2.24$^{***}$ & -0.72 \\
\\[-1.4ex]
& \multicolumn{5}{c}{News Shocks} \\ \cline{2-6} \\[-1.8ex]
IRF Spending: & & & & \\
\quad Small Debt Cost & 0.27$^{*}$ & 1.99$^{***}$ & 2.82$^{***}$ & 2.81$^{***}$ & 1.47$^{**}$ \\
\quad Diff. Costly Debt & 0.00 & -1.37$^{***}$ & -2.30$^{***}$ & -2.38$^{***}$ & -1.18$^{*}$ \\
IRF Output:  & & & & \\
\quad Small Debt Cost & 0.22$^{*}$ & 1.37$^{***}$ & 2.20$^{***}$ & 2.68$^{***}$ & 1.39$^{**}$ \\
\quad Diff. Costly Debt & 0.23 & -0.96$^{***}$ & -1.96$^{***}$ & -2.74$^{***}$ & -1.55$^{**}$ \\
\\[-1.4ex]
& \multicolumn{5}{c}{Timing Res. and News Shocks Combined} \\ \cline{2-6} \\[-1.8ex]
Multipliers: & & & & \\
\quad Small Debt Cost & 1.02$^{***}$ & 0.96$^{***}$ & 0.87$^{***}$ & 0.90$^{***}$ & 0.91$^{***}$ \\
\quad Diff. Costly Debt & -0.37 & -0.42$^{***}$ & -0.23 & -0.22 & -0.22 \\
\\[-1.4ex]
& \multicolumn{5}{c}{Narrative Sign Res. and News Shocks Combined} \\ \cline{2-6} \\[-1.8ex]
Multipliers: & & & & \\
\quad Small Debt Cost & 1.36$^{***}$ & 1.17$^{***}$ & 0.97$^{***}$ & 0.97$^{***}$ & 0.96$^{***}$ \\
\quad Diff. Costly Debt & -0.37 & -0.42$^{***}$ & -0.10 & -0.06 & -0.05 \\
\bottomrule
\end{tabular}

                \label{table main_table_baseline}
                \end{center}
                \begin{flushleft}
                    \begin{small}
                        \textit{Note:}  
                        This table shows the state-dependent impulse response functions (IRF) of output and public spending, for three identification schemes.
                        The two lower panels show the state-dependent multipliers, for two identification schemes. 
                        Small Debt Cost are the point estimates IRF when the service of the debt is initially small. 
                        Diff. Costly Debt are the difference in point estimates IRF between states of large and small debt cost.  
                        The IRF estimates are measured in \% of GDP. The multipliers are measured in dollars. 
                        The forecast horizon ranges between 0 to 16 quarters.    
                        
                        $\star \  p < 0.10$,  $\star \star \  p < 0.05$,  $\star \star \star \ p < 0.01$       
                    \end{small}
                \end{flushleft}
            \end{table}

    \FloatBarrier
    \subsection{The Role of Debt-to-GDP ratio}
    \label{subsection The Role of Debt Quantity}

        Evidence based on panel data suggests that large debt-to-GDP ratios are associated with smaller spending multipliers (see \cite{perottiFiscalPolicyGood1999}, \cite{ilzetzkiHowBigSmall2013}, \cite{nickelFiscalStimulusTimes2014} and \cite{huidromWhyFiscalMultipliers2020}).
        Although not all episodes of costly debt in the sample are driven by large debt-to-GDP ratios (see figure \ref{fig ts_debt_cost_state}), it is critical to disentangle the specific interaction effect of large debt from that of costly debt.

        To test whether it is actually the cost and not the quantity of debt that is driving the results, I augment the models \eqref{eq state dependent lp} and \eqref{eq state dependent multiplier} with an additional state variable. Specifically,  I follow \cite{bernardiniPrivateDebtOverhang2018} and \cite{bronerFiscalMultipliersForeign2022}, rewriting the model \eqref{eq state dependent lp} as
        \begin{equation}
            \label{eq horse race model}
            \begin{array}{cl}
                z_{t+h} = & \left(\alpha_h^A + \beta_h^A shock_t +  \phi_h^A(L) X_{t-1}\right)  +   I_{t-1}^B \left(\alpha_h^B + \beta_h^B shock_t  + \phi_h^B(L) X_{t-1}\right) + \\
                & I_{t-1}^C \left(\alpha_h^C +  \beta_h^C shock_t  + \phi_h^C(L) X_{t-1}\right) + \varepsilon_{t+h}        
            \end{array}
        \end{equation}
        I apply the same transformation to the model \eqref{eq state dependent multiplier}. $ I_{t-1}^B$ ($ I_{t-1}^C $) captures states where the cost (the quantity) of public debt is large. Hence, $\beta_h^A$ is the IRF of $z_{t+h}$ when both the cost and the quantity of debt are small while $\beta_h^A + \beta_h^B$ captures the IRF in states of large cost but small quantity. Conversely,  $\beta_h^A + \beta_h^C$ describes the IRF during episodes when only the debt-to-GDP ratio is large. These states are defined the same way as for the cost of debt, with a comparison to a linear trend.

        Tables \ref{table horse_race_cost_quant_1} and \ref{table horse_race_cost_quant_2} show the IRF of output and spending as well as the multipliers in the baseline states where both the cost and the quantity of debt are small ($\beta_h^A$). In addition, it shows the interaction effect of each state variables ($\beta_h^B$ and $\beta_h^C$). The IRF are estimated for the three identification schemes and the multipliers combine SVAR and news shocks.

        It is clear that the results presented in the previous section are not driven by the effect of large debt-to-GDP ratios. 
        Instead, this exercise reveals even stronger state-dependence in the fiscal cost of public debt. 
        When the debt is initially small but costly, the response of spending is negative after only one year for SVAR shocks and after two years with news shocks. 
        The magnitude of the spending cut is quantitatively important, it peaks down to -2.66\% with timing restriction shocks and to -2.35\% with news shocks. 
        This spending reversal is driving down the response of output, which is negative after two years. 
        The contraction of output peaks down after three years, to -1.28\% with timing restriction shocks.

        Why do we achieve even larger interaction effect of debt cost with this approach? 
        The reason is that the cost and the quantity of the debt are playing in the opposite directions.  
        When the debt is initially high, but associated with a small cost, the response of government spending to the shock is amplified compared to the baseline states.
        With narrative sign restriction shocks, it amounts to 3.14\% after one year and peaks to 3.90\% at two years.  
        This results into a larger response of output, peaking above 3\% for timing and narrative sign restrictions.

        Contrary to the literature, I find that high debt-to-GDP ratios are associated with smaller multipliers only in the short-run. 
        After one year, the interaction effect of debt quantity with the multipliers is no longer significant.  
        Two reasons explain this different findings.
        First, all the previous studies on the relation between debt-to-GDP ratios and fiscal multipliers are based on panel data, with a small time span compared to historical data. 
        Thus, they focus on different variations of debt-to-GDP and government spending than the one analysed here.  
        Secondly, they rely exclusively on SVAR identification. 
        As shown by \cite{rameyIdentifyingGovernmentSpending2011} the resulting shocks can be anticipated, which affects the multipliers.
        Here, the difference in multipliers between states of low and high debt-to-GDP ratios vanishes in the long run as I combine SVAR with news shocks. 
        This suggests that the state-dependence of fiscal multipliers on the quantity of debt might stem from an identification issue with the anticipations.

        The main message from this exercise is that isolating the interaction effect large debt service from the one of large debt-to-GDP ratio reveals even sharper results than those presented in section \ref{subsection Baseline Results}. 
        When the public debt is costly, but its level is low, the initial shock results into spending cut even at medium horizon. This response of the government is self-defeating, as it leads output to contract.

        \begin{table}[h!]
            \centering
            \caption{Difference in IRF: Cost vs Quantity of Debt}
            \vspace{0.2cm} 
                \begin{tabular}{@{\extracolsep{10pt}}lccccc}
\toprule
\bottomrule
\\[-1.4ex]
 & $h = 0$ & $h = 4$ & $h = 8$ & $h=12$ & $h=16$ \\
\midrule
& \multicolumn{5}{c}{Timing Res. Shocks} \\ \cline{2-6} \\[-1.8ex]
IRF Spending: & & & & \\
\quad Baseline States & 0.81$^{***}$ & 1.29$^{***}$ & 1.20$^{***}$ & 1.03$^{***}$ & 0.01 \\
\quad Diff. Cost States & 0.59$^{***}$ & -1.69$^{***}$ & -3.84$^{***}$ & -3.69$^{***}$ & -0.18 \\
\quad Diff. Quantity States & 0.11 & 2.10$^{***}$ & 3.08$^{***}$ & 2.84$^{***}$ & 0.13 \\
IRF Output:  & & & & \\
\quad Baseline States & 1.64$^{***}$ & 2.63$^{***}$ & 2.27$^{***}$ & 1.57$^{***}$ & 0.44 \\
\quad Diff. Cost States & 0.16 & -1.77$^{***}$ & -3.12$^{***}$ & -2.85$^{***}$ & -0.33 \\
\quad Diff. Quantity States & -0.93$^{**}$ & 0.12 & 1.11$^{***}$ & 1.39$^{***}$ & -0.00 \\
\\[-1.4ex]
& \multicolumn{5}{c}{Narrative Sign Res. Shocks} \\ \cline{2-6} \\[-1.8ex]
IRF Spending: & & & & \\
\quad Baseline States & 0.61$^{***}$ & 1.04$^{***}$ & 1.08$^{***}$ & 1.03$^{***}$ & -0.06 \\
\quad Diff. Cost States & 0.45$^{***}$ & -1.44$^{***}$ & -3.43$^{***}$ & -3.31$^{***}$ & 0.05 \\
\quad Diff. Quantity States & 0.38$^{**}$ & 2.10$^{***}$ & 2.82$^{***}$ & 2.51$^{***}$ & 0.09 \\
IRF Output:  & & & & \\
\quad Baseline States & 2.15$^{***}$ & 3.16$^{***}$ & 2.63$^{***}$ & 1.94$^{***}$ & 0.97$^{*}$ \\
\quad Diff. Cost States & 0.33 & -1.57$^{***}$ & -2.81$^{***}$ & -2.48$^{***}$ & 0.01 \\
\quad Diff. Quantity States & -1.42$^{***}$ & -0.47 & 0.52 & 0.78 & -0.70 \\
\\[-1.4ex]
& \multicolumn{5}{c}{News Shocks} \\ \cline{2-6} \\[-1.8ex]
IRF Spending: & & & & \\
\quad Baseline States & 0.71$^{***}$ & 1.87$^{***}$ & 1.29$^{***}$ & 0.23 & -0.59$^{*}$ \\
\quad Diff. Cost States & 0.35 & -1.31$^{***}$ & -2.81$^{***}$ & -3.33$^{***}$ & -2.32$^{***}$ \\
\quad Diff. Quantity States & -0.73$^{**}$ & -0.07 & 1.87$^{***}$ & 3.48$^{***}$ & 2.90$^{***}$ \\
IRF Output:  & & & & \\
\quad Baseline States & 0.61$^{***}$ & 1.17$^{***}$ & 1.32$^{***}$ & 1.06$^{***}$ & 0.06 \\
\quad Diff. Cost States & 0.65$^{***}$ & -0.88$^{***}$ & -2.46$^{***}$ & -3.26$^{***}$ & -2.31$^{***}$ \\
\quad Diff. Quantity States & -0.72$^{***}$ & 0.16 & 1.21$^{***}$ & 1.94$^{***}$ & 1.69$^{**}$ \\
\bottomrule
\end{tabular}

            \label{table horse_race_cost_quant_1}
            \begin{flushleft}
                \begin{small}
                    \textit{Note:}  
                    This table shows the impulse response functions (IRF) of output and public spending with two state variables, for three different identification schemes. 
                    Baseline states are the point estimates IRF when both the cost and the quantity of public debt are small. 
                    Diff. Cost States (Diff. Quantity States) are the difference in point estimates IRF between baseline and states of large debt cost (quantity).
                    The IRF estimates are measured in \% of GDP. 
                    The forecast horizon ranges between 0 to 16 quarters.    
                    
                    $\star \  p < 0.10$,  $\star \star \  p < 0.05$,  $\star \star \star \ p < 0.01$       
                \end{small}
            \end{flushleft}
        \end{table}

        \begin{table}[h!]
            \centering
            \caption{Difference in Multipliers: Cost vs Quantity of Debt}
            \vspace{0.2cm} 
                \begin{tabular}{@{\extracolsep{10pt}}lccccc}
\toprule
\bottomrule
\\[-1.4ex]
 & $h = 0$ & $h = 4$ & $h = 8$ & $h=12$ & $h=16$ \\
\midrule
& \multicolumn{5}{c}{Timing Res. and News Shocks Combined} \\ \cline{2-6} \\[-1.8ex]
Multipliers: & & & & \\
\quad Baseline States & 2.13$^{***}$ & 1.30$^{***}$ & 0.80$^{***}$ & 1.08$^{***}$ & 1.20$^{***}$ \\
\quad Diff. Cost States & -0.04 & -0.32$^{**}$ & -0.29$^{*}$ & -0.31$^{*}$ & -0.33 \\
\quad Diff. Quantity States & -1.49$^{***}$ & -0.40$^{*}$ & 0.09 & -0.23 & -0.38 \\
\\[-1.4ex]
& \multicolumn{5}{c}{Narrative Sign Res. and News Shocks Combined} \\ \cline{2-6} \\[-1.8ex]
Multipliers: & & & & \\
\quad Baseline States & 3.24$^{***}$ & 1.44$^{***}$ & 0.68$^{***}$ & 1.01$^{***}$ & 1.12$^{***}$ \\
\quad Diff. Cost States & 0.28 & -0.29 & -0.26 & -0.27 & -0.28 \\
\quad Diff. Quantity States & -2.67$^{***}$ & -0.48$^{*}$ & 0.27$^{*}$ & -0.12 & -0.25 \\
\bottomrule
\end{tabular}

            \label{table horse_race_cost_quant_2}
            \begin{flushleft}
                \begin{small}
                    \textit{Note:}  
                    This table shows the cumulative multipliers with two state variables, for two different identification schemes. 
                    Baseline states are the point estimates of multipliers when both the cost and the quantity of public debt are small. 
                    Diff. Cost States (Diff. Quantity States) are the difference in point estimates of multipliers between baseline and states of large debt cost (quantity).
                    The multipliers estimates are measured in dollars.
                    The forecast horizon ranges between 0 to 16 quarters.    
                    
                    $\star \  p < 0.10$,  $\star \star \  p < 0.05$,  $\star \star \star \ p < 0.01$       
                \end{small}
            \end{flushleft}
        \end{table}

    \subsection{The Dynamic of the Government Budget Constraint}
    \label{subsection The Dynamic of the Government Budget Constraint}

        I now investigate the mechanisms underlying the government's asymmetric response to spending shocks.
        The findings of the previous section show that if a spending shock hits the economy when servicing the public debt induces a large fiscal cost, the response of public spending is limited and short-lived. 
        It is thus natural  to expect asymmetric adjustments along the other dimensions of the government budget constraint in response to the shock. 
        To test this hypothesis, I employ the baseline model \eqref{eq state dependent lp} to evaluate the state-dependent IRF for the different budget constraint components, that is public spending, tax and debt. 
        Table \ref{table bc_irf} shows the resulting estimates for the three identification schemes used in the paper.

        Depending on the initial cost of debt, the IRF of government tax and debt vary significantly.
        A shock hitting the economy during periods of low debt cost triggers a continuous increase in public debt, amounting to 4.15\% after four years with timing restriction identification. 
        Although we do observe a positive response of tax revenues at medium and long horizon, the magnitude of this increase is limited compared to the IRF for public spending. 
        This points out that the bulk of the additional spending are financed through debt rather than by tax, which is consistent with the nature of fiscal stimulus generally analysed in the literature. 
        When the cost of debt is initially high, the response of tax is even more limited.
        With timing and narrative sign restriction identification, the effect of the shock on tax is close to 0. 
        With news shocks, it is positive only at short horizon. 
        The pattern of the IRF for public debt is also strongly affected compared to the low debt cost state. 
        Here, for the three identification schemes, the long term response of public debt is close to 0. 
        This is in sharp contrast with the large debt build-up observed when the debt cost is initially small. 
        Overall, we observe two very different effects of an initial spending shock on the government budget constraint depending on the initial cost of debt.
        When this cost is limited, the government increases spending, which is financed mostly through debt build-up and partly through more tax revenues. 
        By contrast if the cost of debt is large prior to the shock, the government limits the debt issuance in the short run to keep constant the stock of public debt in the long run.
        As tax revenues are also constant, the initial increase is compensated by spending reversal at medium horizon.

        These results suggest that the fiscal authority sacrifices public spending to debt sustainability when the initial cost of debt is large. 
        Yet, it is a priori unclear whether this reaction is the result of the fiscal authority itself or whether it is shaped by a constraint exerted by the monetary authority or the financial markets. 
        To address this question, I estimate the state-dependent IRF of the aggregate debt ceiling\footnote{Before 1939, there were no aggregate limit on the federal debt. Instead, Congress imposed limits on every security. \cite{hallHistoryDebtLimits2015} use these security level limits to infer the aggregate debt limit since 1776. I use the same series in this analysis.}, the federal fund rate and the 10 year Treasury yield to a spending shock. I augment the model \eqref{eq state dependent lp} by using lagged values of the dependent variable, and lagged values of the inflation rate when evaluating the response of the federal fund rate and the market yield. Table \ref{table discussion_table} reports the results.

        The IRF of the aggregate debt ceiling suggests that the asymmetric response of public debt is the own decision of the fiscal authority. 
        During episodes of small debt cost, the response of the debt ceiling to spending shock is monotonic and follows closely the response of public debt. 
        Yet, the debt ceiling increases by a lesser extent than the public debt. 
        That is, as Congress allows for further debt financing, it reduces fiscal space. 
        This suggests that the fiscal authority preserves a concern for debt sustainability in these states, while being able to finance public spending through debt, due to favourable initial fiscal conditions. 
        The response of the fiscal authority differs sharply when the cost of debt is high prior to the shock. 
        At long horizon, the aggregate limit on public debt is reduced, for the three different identification strategies. 
        With the shocks identified through SVAR, we observe a slight increase at short horizon.  
        When using news shocks, the constraint on public debt during costly episodes appears even more clearly.  
        Even at short horizons, Congress lowers the aggregate debt ceiling. 
        In sum, when public debt is initially costly, Congress decides to tie its own hands on fiscal policy by tightening the constraint on public debt. 
        This suggests that in these states, the fiscal authority prefers to hold the stock of public debt constant in response to the shock, even if this requires reducing the response of public spending.

        Eventually, I discuss the potential role of the FED and the financial markets in constraining the fiscal response of the government to spending shocks.
        For the FED, I evaluate the state-dependent response of the federal fund rate (which reduces the sample to the earliest available date of observation, that is 1928, see \cite{anbilNewDailyFederal2021}). 
        This exercise suggests that the response of the FED to a spending shock during times of small debt cost is slightly expansionary. 
        For the three identification strategy, we observe a reduction in the fed fund rate at medium horizon.
        Yet, this decline is small in magnitude, it amounts at most at -22 basis points, two years after a timing restriction shock. 
        By contrast, with large debt cost, the response of the fed fund rate is essentially zero.
        Therefore, there is little evidence is attempting to coerce the fiscal authority on its policy during periods of adverse fiscal conditions. 
        At best, it is not as accommodative as it is during episodes of small debt cost.
     
        To gauge the response of the financial market, I estimate the state-dependent IRF of the 10 year Treasury yield. 
        Here, the three different identification schemes result in mixed evidence. 
        For the shocks identified with structural restrictions, the response of the 10 year yield is positive and significant in states of small debt cost, while it is negative in states of large debt cost. 
        By contrast, when using news shocks, the market yield does not respond to spending shocks in states of small debt cost, but it increases at medium horizon if the shock hits in times of high debt cost. 
        Shocks identified through SVAR can be anticipated.
        Thus, to evaluate a market response, the results obtained with news shocks can be given more credence. 
        This line of results suggests that the financial market could play a role in constraining the fiscal policy when the cost of debt is large. 
        However, the magnitude of this constraint is small, the peak increase in the market yield is 10 basis point after two years. 
        Moreover, after four years the response of the financial markets is no longer significant, while most of the contraction in public debt and public spending occurs at long horizon.

        Overall, there is only mixed evidence that the financial markets or the FED is constraining the response of the fiscal authority. 
        By contrast, it appears clearly that the fiscal authority is tying its own hands when a spending shock hits while the debt is costly. 
        It does so by reducing the aggregate debt ceiling, which constrains its borrowing capacity and hence its ability to finance further government spending. 
        This suggests that under adverse fiscal conditions, the government is willing to sacrifice the response of public spending in order to preserve the sustainability of the public debt.

        \begin{table}[h!]
            \centering
            \caption{Budget Constraint Decomposition}
            \vspace{0.2cm} 
                \begin{tabular}{@{\extracolsep{10pt}}lccccc}
\toprule
\bottomrule
\\[-1.4ex]
 & $h = 0$ & $h = 4$ & $h = 8$ & $h=12$ & $h=16$ \\
\midrule
& \multicolumn{5}{c}{Timing Res. Shocks} \\ \cline{2-6} \\[-1.8ex]
IRF Spending: & & & & \\
\quad Small Debt Cost & 1.09$^{***}$ & 2.22$^{***}$ & 2.71$^{***}$ & 2.36$^{***}$ & 0.24 \\
\quad Diff. Costly Debt & 0.50$^{***}$ & -0.71$^{***}$ & -2.51$^{***}$ & -2.68$^{***}$ & -0.26 \\
IRF Debt:  & & & & \\
\quad Small Debt Cost & 0.06 & 1.44$^{***}$ & 2.92$^{***}$ & 4.13$^{***}$ & 4.15$^{***}$ \\
\quad Diff. Costly Debt & -0.02 & 0.85$^{**}$ & -1.06$^{*}$ & -3.22$^{***}$ & -3.84$^{***}$ \\
IRF Tax: & & & & \\
\quad Small Debt Cost & 0.15$^{*}$ & 0.68$^{***}$ & 0.91$^{***}$ & 0.80$^{***}$ & 0.36$^{**}$ \\
\quad Diff. Costly Debt & 0.04 & -0.64$^{***}$ & -0.96$^{***}$ & -0.83$^{***}$ & -0.32$^{*}$ \\
\\[-1.4ex]
& \multicolumn{5}{c}{Narrative Sign Res. Shocks} \\ \cline{2-6} \\[-1.8ex]
IRF Spending: & & & & \\
\quad Small Debt Cost & 1.00$^{***}$ & 2.00$^{***}$ & 2.49$^{***}$ & 2.21$^{***}$ & 0.09 \\
\quad Diff. Costly Debt & 0.47$^{***}$ & -0.55$^{**}$ & -2.31$^{***}$ & -2.51$^{***}$ & -0.05 \\
IRF Debt:  & & & & \\
\quad Small Debt Cost & 0.08 & 1.30$^{***}$ & 2.64$^{***}$ & 3.76$^{***}$ & 3.72$^{***}$ \\
\quad Diff. Costly Debt & -0.00 & 0.91$^{***}$ & -0.82 & -2.86$^{***}$ & -3.37$^{***}$ \\
IRF Tax: & & & & \\
\quad Small Debt Cost & 0.12$^{*}$ & 0.63$^{***}$ & 0.83$^{***}$ & 0.73$^{***}$ & 0.30$^{**}$ \\
\quad Diff. Costly Debt & 0.04 & -0.60$^{***}$ & -0.89$^{***}$ & -0.76$^{***}$ & -0.25 \\
\\[-1.4ex]
& \multicolumn{5}{c}{News Shocks} \\ \cline{2-6} \\[-1.8ex]
IRF Spending: & & & & \\
\quad Small Debt Cost & 0.27$^{*}$ & 1.99$^{***}$ & 2.82$^{***}$ & 2.81$^{***}$ & 1.47$^{**}$ \\
\quad Diff. Costly Debt & 0.00 & -1.37$^{***}$ & -2.30$^{***}$ & -2.38$^{***}$ & -1.18$^{*}$ \\
IRF Debt:  & & & & \\
\quad Small Debt Cost & 0.02 & 1.47$^{***}$ & 3.43$^{***}$ & 5.12$^{***}$ & 6.14$^{***}$ \\
\quad Diff. Costly Debt & -0.37$^{**}$ & -1.82$^{***}$ & -3.97$^{***}$ & -5.73$^{***}$ & -6.82$^{***}$ \\
IRF Tax: & & & & \\
\quad Small Debt Cost & 0.00 & 0.51$^{***}$ & 0.91$^{***}$ & 0.99$^{***}$ & 0.70$^{***}$ \\
\quad Diff. Costly Debt & 0.49$^{***}$ & -0.16$^{*}$ & -0.65$^{***}$ & -0.85$^{***}$ & -0.73$^{***}$ \\
\bottomrule
\end{tabular}

            \label{table bc_irf}
            \begin{flushleft}
                \begin{small}
                    \textit{Note:}  
                    This table shows the state-dependent impulse response functions (IRF) of public spending, debt and taxes, for three identification schemes.
                    Small Debt Cost are the point estimates IRF when the service of the debt is initially small. 
                    Diff. Costly Debt are the difference in point estimates IRF between states of large and small debt cost.  
                    The IRF estimates are measured in \% of GDP. 
                    The forecast horizon ranges between 0 to 16 quarters.    
                    
                    $\star \  p < 0.10$,  $\star \star \  p < 0.05$,  $\star \star \star \ p < 0.01$   
                \end{small}
            \end{flushleft}
        \end{table}

        \begin{table}[h!]
            \centering
            \caption{Responses of Markets, Fiscal and Monetary Authority}
            \vspace{0.2cm} 
                \begin{tabular}{@{\extracolsep{10pt}}lccccc}
\toprule
\bottomrule
\\[-1.4ex]
 & $h = 0$ & $h = 4$ & $h = 8$ & $h=12$ & $h=16$ \\
\midrule
& \multicolumn{5}{c}{Timing Res. Shocks} \\ \cline{2-6} \\[-1.8ex]
IRF Debt Ceiling:  & & & & \\
\quad Small Debt Cost & -0.24 & 0.77$^{*}$ & 1.91$^{***}$ & 3.15$^{***}$ & 4.36$^{***}$ \\
\quad Diff. Costly Debt & -0.47 & 0.57 & -0.51 & -2.62$^{***}$ & -4.53$^{***}$ \\
IRF FFR: & & & & \\
\quad Small Debt Cost & -0.05 & -0.16$^{**}$ & -0.22$^{***}$ & -0.18$^{**}$ & -0.11 \\
\quad Diff. Costly Debt & -0.06 & -0.03 & 0.12 & 0.14 & 0.05 \\
IRF 10 year yield:  & & & & \\
\quad Small Debt Cost & -0.01 & 0.03$^{*}$ & 0.07$^{***}$ & 0.08$^{***}$ & 0.08$^{*}$ \\
\quad Diff. Costly Debt & 0.02 & -0.10$^{***}$ & -0.16$^{***}$ & -0.13$^{***}$ & -0.09$^{*}$ \\
\\[-1.4ex]
& \multicolumn{5}{c}{Narrative Sign Res. Shocks} \\ \cline{2-6} \\[-1.8ex]
IRF Debt Ceiling:  & & & & \\
\quad Small Debt Cost & -0.31 & 0.38 & 1.46$^{***}$ & 2.70$^{***}$ & 3.79$^{***}$ \\
\quad Diff. Costly Debt & -0.54 & 0.86$^{*}$ & -0.10 & -2.20$^{***}$ & -3.93$^{***}$ \\
IRF FFR: & & & & \\
\quad Small Debt Cost & -0.04 & -0.13$^{**}$ & -0.20$^{***}$ & -0.17$^{**}$ & -0.10 \\
\quad Diff. Costly Debt & -0.06 & -0.06 & 0.08 & 0.11 & 0.02 \\
IRF 10 year yield:  & & & & \\
\quad Small Debt Cost & -0.01 & 0.01 & 0.07$^{***}$ & 0.08$^{**}$ & 0.10$^{*}$ \\
\quad Diff. Costly Debt & 0.00 & -0.09$^{***}$ & -0.16$^{***}$ & -0.12$^{***}$ & -0.12$^{*}$ \\
\\[-1.4ex]
& \multicolumn{5}{c}{News Shocks} \\ \cline{2-6} \\[-1.8ex]
IRF Debt Ceiling:  & & & & \\
\quad Small Debt Cost & -0.16 & 1.81$^{***}$ & 3.18$^{***}$ & 4.21$^{***}$ & 5.74$^{***}$ \\
\quad Diff. Costly Debt & -0.35 & -2.65$^{***}$ & -4.36$^{***}$ & -5.31$^{***}$ & -6.64$^{***}$ \\
IRF FFR: & & & & \\
\quad Small Debt Cost & -0.04 & -0.13$^{***}$ & -0.16$^{***}$ & -0.12$^{**}$ & -0.10 \\
\quad Diff. Costly Debt & -0.03 & 0.18$^{**}$ & 0.26$^{***}$ & 0.11 & 0.08 \\
IRF 10 year yield:  & & & & \\
\quad Small Debt Cost & -0.01 & -0.02$^{**}$ & -0.01 & 0.01 & -0.01 \\
\quad Diff. Costly Debt & 0.02 & 0.09$^{***}$ & 0.10$^{***}$ & 0.06$^{***}$ & 0.04 \\
\bottomrule
\end{tabular}

            \label{table discussion_table}
            \begin{flushleft}
                \begin{small}
                    \textit{Note:}  
                    This table shows the state-dependent impulse response functions (IRF) of the debt ceiling, the federal fund rate and the market yield on 10 years Treasury, for three identification schemes.
                    Small Debt Cost are the point estimates IRF when the service of the debt is initially small. 
                    Diff. Costly Debt are the difference in point estimates IRF between states of large and small debt cost.  
                    The estimates are measured in \% of GDP. 
                    The forecast horizon ranges between 0 to 16 quarters.    
                    
                    $\star \  p < 0.10$,  $\star \star \  p < 0.05$,  $\star \star \star \ p < 0.01$   
                \end{small}
            \end{flushleft}
        \end{table}

\section{Sensitivity Analysis}
\label{section Sensitivity Analysis}

    \subsection{Continuous State Variable}
    \label{subsection continuous state variable}

        Allowing for only two potential states of the economy provides a clear interpretation of the state-dependent IRF, as the cost of public debt is either below or above its trend when a shock occurs.
        This binary definition is also likely to coincide with the model of the policymaker, who chooses its response according to two possible regimes, whether servicing the debt creates fiscal pressure or not.

        However, as explained in section \ref{subsection Data Description and state variable}, this approach incurs a parametric cost, as it requires taking a stand on the definition of the states. 
        In the appendix \ref{appendix section Alternative states definition}, I show that the findings of the paper are robust to variations in these parameters. 
        To further assess the robustness, I adopt an alternative approach following \cite{bronerFiscalMultipliersForeign2022} and consider a continuous state variable. 
        Using the fiscal cost of public debt directly as a state variable saves on the parameters as no 
        additional assumptions are required to estimate state-dependent impulse responses. The model for the IRF of output and spending writes : 
        \begin{equation}
                z_{t+h} =  \left(\alpha_h^A + \beta_h^A shock_t + \Phi_h^A(L) X_{t-1}\right) +  \text{fiscal cost}_{t-1} \left(\alpha_h^B + \beta_h^B shock_t + \Phi_h^B(L) X_{t-1}   \right) + \varepsilon_{t+h}
        \end{equation}
        
        The model \eqref{eq state dependent multiplier} is adapted in the same way to estimate the multipliers with a continuous state variable. 
        With this approach, $\beta_h^A$ captures the response to the shock, in the out-of-sample case where the cost of debt prior to the shock is equal to 0. 
        The coefficient $\beta_h^B$ is the interaction of spending shocks with the cost of debt. 
        Thus, the state-dependent response is given by $ \beta_h^A +  \beta_h^B. \text{fiscal cost}_{t-1}$, and is defined for a given level of fiscal cost of public debt. 
        This highlights a limitation of this approach, since the states are not defined, one needs to choose arbitrary levels for the cost of debt to compute the state-dependent responses. 

        \begin{figure}[h!]
            \caption{State-Dependent Responses: Continuous State}
            \begin{center}
                \includegraphics{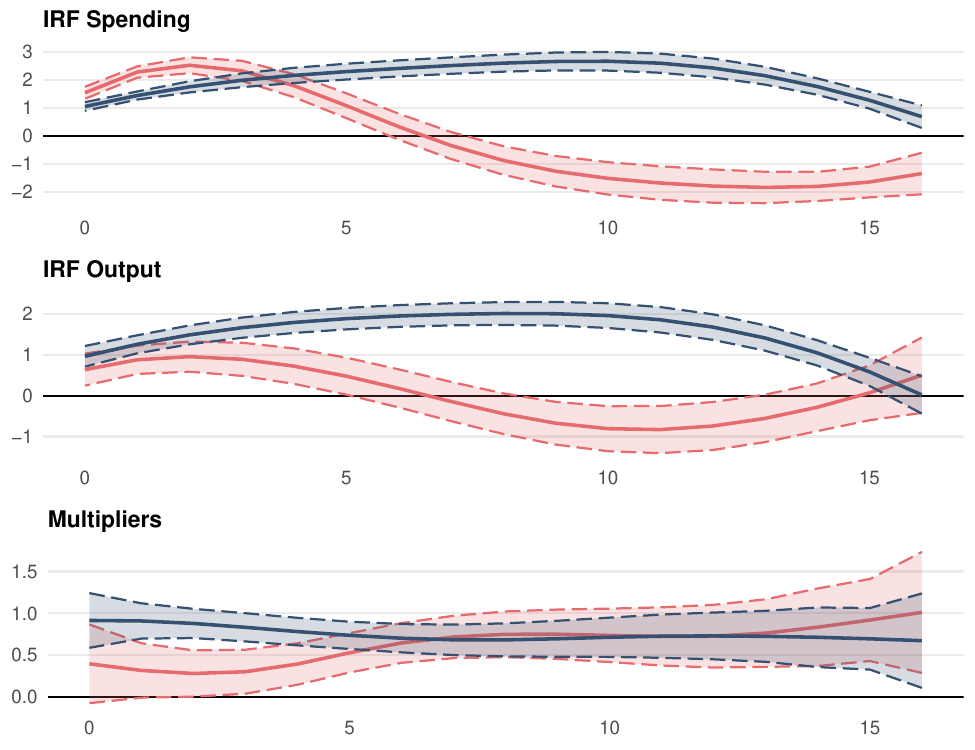} \label{fig continuous_states_irf}
            \end{center}
            \vspace{-0.6cm}
            \begin{small}
                \textit{Note:}  
                This figure shows the state-dependent impulse response functions (IRF) to a government spending shock identified with a timing restriction. 
                The states are defined with a continuous variable.
                The IRF in red (blue) corresponds to a cost of public debt set to its 90th (10th) percentile. 
                The two upper panels show the IRF of public spending and output, measured in \% of potential GDP. 
                The last panel shows the cumulative multiplier, measured in dollars.
                The IRF are estimated using smooth local projections.
                The time horizon is in quarters. 
                The confidence intervals are defined at a 10\% risk level.  
            \end{small}
        \end{figure}

        Figure \ref{fig continuous_states_irf} shows the state-dependent IRF and multipliers obtained with this approach, setting the value of debt cost equal to its 10th and 90th percentile. 
        It suggests that relaxing assumptions on the definition of states does not change the main findings presented in section \ref{subsection Baseline Results}. 
        The multiplier is smaller at short-horizon when the cost of debt is initially high. 
        The IRF of output and spending display strong state-dependence.
        The asymmetry between cases of large and small debt service is even more pronounced than in the baseline results.
        The reason is that we consider here more extreme values for the initial states.
        By contrast, when defining a dummy state variable, one pools together episodes with small and large deviation from the time trend, which dampens the interaction effect. 
        Thus, it is worthwhile noting that for larger values of debt cost in the sample, the spending reversal that occurs two years after the shock is more pronounced.
        The IRF of spending peaks down close to -2\%. 
        In this situation, the fiscal policy is self-defeating in the long-run, with an output IRF peaking down near -1\%  after 11 quarters.

    \subsection{Confounding State Variables}
    \label{subsection confounding state variables}

        The results of section \ref{subsection Baseline Results} could be spuriously driven by other changes in the state of the economy that are concomitant with episodes of costly debt.  
        In particular, evidence from \cite{rameyGovernmentSpendingMultipliers2018} suggests that episodes of economic slack are associated with larger IRF for output and spending (although the multipliers are identic). 
        Thus, it is a concern that the more aggressive response of the government observed during episodes of large debt cost may be due to coincidental negative output gap. 
        Similarly, the results could be confounded by states of Zero Lower Bound, as they mechanically push down the fiscal cost of public debt.

        To address these concerns, I employ the same model as in section \ref{subsection The Role of Debt Quantity} to run two separate "horse races" between the cost of public debt and ZLB or slack states. 
        I follow \cite{rameyGovernmentSpendingMultipliers2018} and define states of slack as episodes where the unemployment rate is greater than 6.5\% and states of ZLB as the periods 1932Q2-1951Q1 and 2008Q4-2015Q4.
        Table \ref{table horse_race_zlb_slack} shows the resulting estimates for the IRF and multipliers in the baseline states, as well as the interaction term of each state variable. 
        These baseline states are defined as episodes where the cost of debt is small and the economy is away from either the ZLB or the slack. 
        The IRF are estimated using timing restriction shocks and the multipliers by combining timing restriction with news shocks. 

        The findings of section \ref{subsection Baseline Results} are still valid under this specification. 
        When controlling for the interaction effect of ZLB states, we achieve very similar results to the main ones.
        Yet, it can be noted that episodes of ZLB are associated with larger spending IRF. 
        Therefore, muting this interaction effect magnifies the decline in government spending that occurs in response to shocks hitting in states of costly debt. 
        After two years, the costly debt IRF for spending is -0.80\%, against 0.20\% with the main specification.  

        Considering states of economic slack has little effect on the estimates.
        After two years, a shock that occurred during episodes of large debt service results into a 0.12\% increase in government spending and a 0.28\% increase in output, compared to respectively 0.20\% and 0.24\% in the main results of section \ref{subsection Baseline Results}.    
        Episodes of high unemployment are associated with a positive and large interaction effect. 
        Yet, controlling for this effect mostly affects the estimates for episodes of small debt cost.  
        Thus, a part of the large IRF reported in section \ref{subsection Baseline Results} for states of small debt cost is actually driven by a concurrent  dependence in states of slack. 
        Importantly, this does not change the main estimates in states of costly public debt, which are the key findings of this analysis.

        \begin{table}[h!]
            \centering
            \caption{Horse Race: ZLB and Slack States}
            \vspace{0.2cm} 
                \begin{tabular}{@{\extracolsep{10pt}}lccccc}
\toprule
\bottomrule
\\[-1.4ex]
 & $h = 0$ & $h = 4$ & $h = 8$ & $h=12$ & $h=16$ \\
\midrule
& \multicolumn{5}{c}{Horse Race ZLB} \\ \cline{2-6} \\[-1.8ex]
IRF Spending: & & & & \\
\quad Baseline States & 0.94$^{***}$ & 1.62$^{***}$ & 1.87$^{***}$ & 1.64$^{***}$ & 0.10 \\
\quad Diff. Cost States & 0.62$^{***}$ & -0.93$^{***}$ & -2.67$^{***}$ & -2.63$^{***}$ & 0.17 \\
\quad Diff. ZLB States & -0.02 & 0.48$^{**}$ & 1.05$^{***}$ & 0.90$^{***}$ & -0.92 \\
IRF Output:  & & & & \\
\quad Baseline States & 1.04$^{***}$ & 2.03$^{***}$ & 2.28$^{***}$ & 1.96$^{***}$ & 0.34 \\
\quad Diff. Cost States & -0.13 & -1.57$^{***}$ & -2.86$^{***}$ & -2.57$^{***}$ & 0.04 \\
\quad Diff. ZLB States & 0.11 & 0.45$^{*}$ & 0.58$^{**}$ & 0.41 & -0.59 \\
Multipliers: & & & & \\
\quad Baseline States & 1.12$^{***}$ & 1.05$^{***}$ & 0.94$^{***}$ & 1.13$^{***}$ & 1.24$^{***}$ \\
\quad Diff. Cost States & -0.33 & -0.41$^{***}$ & -0.20 & -0.26$^{*}$ & -0.27 \\
\quad Diff. ZLB States & -0.02 & 0.11 & 0.12 & -0.15 & -0.31 \\
\\[-1.4ex]
& \multicolumn{5}{c}{Horse Race Slack} \\ \cline{2-6} \\[-1.8ex]
IRF Spending: & & & & \\
\quad Baseline States & 1.14$^{***}$ & 1.66$^{***}$ & 1.49$^{***}$ & 1.10$^{***}$ & -0.28 \\
\quad Diff. Cost States & 0.44$^{**}$ & -0.09 & -1.37$^{***}$ & -1.43$^{***}$ & 0.16 \\
\quad Diff. Slack States & -0.35$^{**}$ & 0.57$^{**}$ & 1.69$^{***}$ & 1.92$^{***}$ & 0.65 \\
IRF Output:  & & & & \\
\quad Baseline States & 0.96$^{***}$ & 1.16$^{***}$ & 0.97$^{***}$ & 0.85$^{***}$ & 0.32 \\
\quad Diff. Cost States & -0.01 & -0.12 & -0.69$^{***}$ & -0.83$^{***}$ & -0.24 \\
\quad Diff. Slack States & 0.19 & 1.66$^{***}$ & 2.31$^{***}$ & 2.44$^{***}$ & 1.51$^{*}$ \\
Multipliers: & & & & \\
\quad Baseline States & 0.93$^{***}$ & 0.61$^{***}$ & 0.50$^{***}$ & 0.62$^{***}$ & 0.68$^{***}$ \\
\quad Diff. Cost States & -0.20 & 0.02 & 0.20 & 0.06 & -0.01 \\
\quad Diff. Slack States & 0.15 & 0.65$^{***}$ & 0.54$^{***}$ & 0.36$^{**}$ & 0.32 \\
\bottomrule
\end{tabular}

            \label{table horse_race_zlb_slack}
            \begin{flushleft}
                \begin{small}
                    \textit{Note:}  
                    This table shows the impulse response functions (IRF) of output and public spending as well as the cumulative multipliers, with two state variables. 
                    Baseline states are the point estimates when the cost of debt is small and the economy is off the states of zero lower bound or of slack. 
                    Diff. Cost States (Diff. ZLB States) [Diff. Slack States] are the difference in point estimates between baseline and states of large debt cost (ZLB) [Slack].
                    The IRF estimates are measured in \% of GDP. The multipliers are measured in dollars. 
                    The forecast horizon ranges between 0 to 16 quarters. 
                    
                    $\star \  p < 0.10$,  $\star \star \  p < 0.05$,  $\star \star \star \ p < 0.01$   
                \end{small}
            \end{flushleft}
        \end{table}

\FloatBarrier
\section{Conclusions}
\label{section Conclusions}

        This paper explores how the service of the debt affects fiscal policy and its effect on the economy. 
        Using U.S. historical data, I evaluate the response of government spending and output to spending shocks, allowing for state-dependence in the cost of public debt. 
        Instead of considering an exogenous process for fiscal policy, I assume spending shocks to be changes in the preference for public spending.
        This perspective allows for an endogenous response of public spending, which can vary with the initial fiscal conditions. 

        The main findings of the paper are as follows. 
        The response of government spending is strongly state-dependent with the service of the debt.
        When the cost of debt is large, the response of government spending quickly reverts to zero. 
        The response of output closely follows the shape of fiscal policy. 
        When the service of the debt is burdening, spending shocks are having a moderate and short-lived effect on output. 
        Importantly, the asymmetry of fiscal policy is more important than the multipliers in explaining the difference in output response.
        Eventually, the shape of fiscal policy is mostly determined by the fiscal authority itself, rather than by market or monetary authority constraints. 
        
        These findings have implications for the discussion on public debt sustainability. 
        The evidence suggests that historically, the fiscal authority has been willing to sacrifice public spending when the sustainability of the debt was at risk. 
        Although this policy maintains the public debt on a sustainable path, it costs the government its ability to stimulate output when the preference for public spending increases. 
        
        These results also inform the literature on fiscal policy, which focuses on estimating the multipliers.
        This paper highlights the importance of analysing the response of government spending as well.
        This requires to think of fiscal policy as endogenous and hence to change perspective on spending shocks. 
        In this paper, I propose to consider them as shocks to the preference for public spending, instead of shocks to the quantity of public spending, but further research could be undertaken to developing alternative views.

\FloatBarrier
\FloatBarrier
\printbibliography

\appendix

\newpage
\FloatBarrier

\FloatBarrier
\section{Smooth Local Projection vs Local Projection}
\label{appendix section Smooth Local Projection vs Local Projection}

    This section is to assess the robustness of the results to using smooth local projections (\cite{barnichonImpulseResponseEstimation2019}) for the estimation of the impulse response functions and the cumulative multipliers. 

    \FloatBarrier
    \subsection{Degree of Shrinkage}
    \label{appendix subsection Degree of Shrinkage}
        
        The baseline results are obtained using a degree of shrinkage $r=3$. This amounts to shrink the impulse response functions toward a second order polynomial function of the forecast horizon. 
        This choice is motivated by the common inverted U-shape of impulse response functions, so as to introduce the minimal bias compared to standard local projections.
        As shown in table \ref{table appendix_degree_shrinkage.tex}, adopting another degree of shrinkage does not sensibly modifies the results obtained with the baseline specification.

    \subsection{Standard Local Projections}
    \label{appendix subsection Standard Local Projection}

\begin{figure}[h!]
          \caption{State-Dependent Responses: Standard Local Projections}
          \begin{center}
            \includegraphics{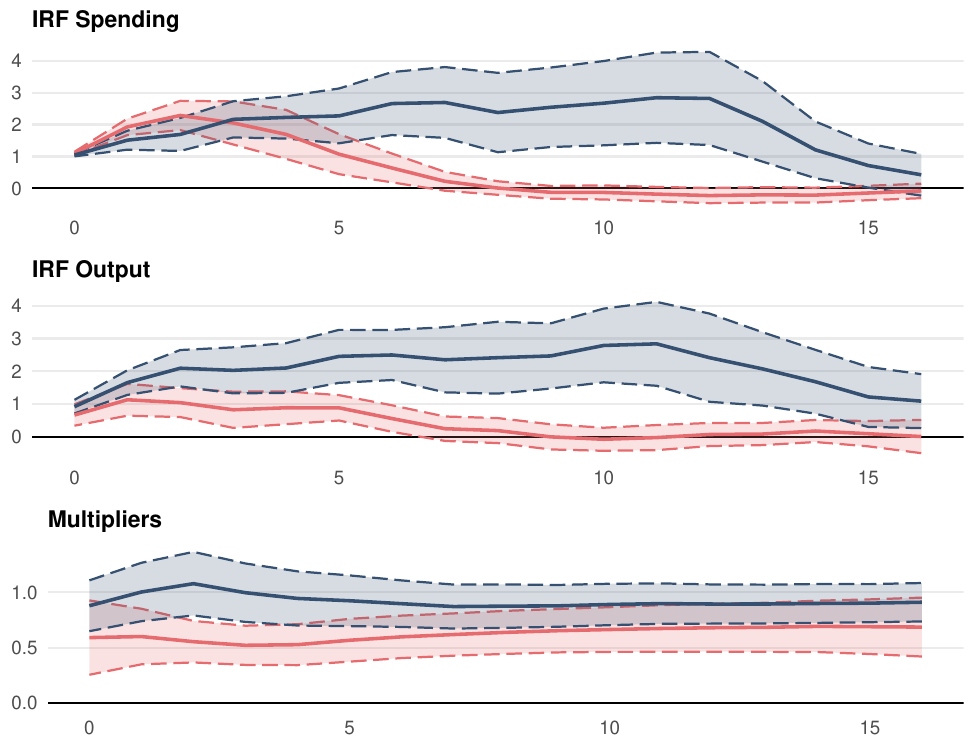} \label{fig appendix_standard_lp}
          \end{center}
          \vspace{-0.6cm}
          \begin{small}
            \textit{Note:}  his figure shows the state-dependent impulse response functions (IRF) to a government spending shock identified with a timing restriction. 
            The IRF in red (blue) corresponds to initially large (small) debt cost.
            The two upper panels show the IRF of public spending and output, measured in \% of potential GDP. 
            The last panel shows the cumulative multiplier, measured in dollars.
            The IRF are estimated using standard local projections à la \cite{jordaEstimationInferenceImpulse2005}.
            The time horizon is in quarters. 
            The confidence intervals are defined at a 10\% risk level.  
          \end{small}
        \end{figure}

        To ensure that adopting smooth local projections does not generate spurious results, I also estimate the impulse response functions and multipliers using standard local projections à la \cite{jordaEstimationInferenceImpulse2005}.
        As shown in figure \ref{fig appendix_standard_lp}, the shape of the IRF and multipliers estimated with standard local projections closely follows the ones obtained with smooth local projections. 
        This confirms that using smooth local projections only introduce a minor bias compared to \cite{jordaEstimationInferenceImpulse2005} approach. 
        Yet, it appears clearly that standard local projections result into larger standard errors. 
        Smooth local projections provide an efficient way to reduce the noise around the IRF estimates.

        \begin{table}[h!]
            \centering
            \caption{Robustness: Degree of Shrinkage}
            \vspace{0.2cm} 
                \begin{tabular}{@{\extracolsep{10pt}}lccccc}
\toprule
\bottomrule
\\[-1.4ex]
 & $h = 0$ & $h = 4$ & $h = 8$ & $h=12$ & $h=16$ \\
\midrule
& \multicolumn{5}{c}{Degree of Shrinkage $r = 1$} \\ \cline{2-6} \\[-1.8ex]
IRF Spending: & & & & \\
\quad Baseline States & 1.95$^{***}$ & 2.06$^{***}$ & 2.11$^{***}$ & 1.99$^{***}$ & 1.77$^{***}$ \\
\quad Diff. Costly Debt & -0.96$^{***}$ & -1.20$^{***}$ & -1.59$^{***}$ & -1.70$^{***}$ & -1.56$^{***}$ \\
IRF Output:  & & & & \\
\quad Baseline States & 1.97$^{***}$ & 2.08$^{***}$ & 2.15$^{***}$ & 2.08$^{***}$ & 1.94$^{***}$ \\
\quad Diff. Costly Debt & -1.39$^{***}$ & -1.55$^{***}$ & -1.77$^{***}$ & -1.81$^{***}$ & -1.71$^{***}$ \\
Multipliers : & & & & \\
\quad Baseline States & 1.01$^{***}$ & 0.95$^{***}$ & 0.87$^{***}$ & 0.90$^{***}$ & 0.91$^{***}$ \\
\quad Diff. Costly Debt & -0.45$^{***}$ & -0.41$^{***}$ & -0.24 & -0.21 & -0.22 \\
\\[-1.4ex]
& \multicolumn{5}{c}{Degree of Shrinkage $r = 2$} \\ \cline{2-6} \\[-1.8ex]
IRF Spending: & & & & \\
\quad Baseline States & 1.31$^{***}$ & 2.21$^{***}$ & 2.62$^{***}$ & 2.15$^{***}$ & 0.78$^{*}$ \\
\quad Diff. Costly Debt & 0.63$^{***}$ & -0.96$^{***}$ & -2.28$^{***}$ & -2.31$^{***}$ & -1.09$^{**}$ \\
IRF Output:  & & & & \\
\quad Baseline States & 1.30$^{***}$ & 2.20$^{***}$ & 2.58$^{***}$ & 2.27$^{***}$ & 1.18$^{***}$ \\
\quad Diff. Costly Debt & -0.31 & -1.45$^{***}$ & -2.31$^{***}$ & -2.22$^{***}$ & -1.15$^{**}$ \\
Multipliers : & & & & \\
\quad Baseline States & 1.07$^{***}$ & 0.96$^{***}$ & 0.87$^{***}$ & 0.90$^{***}$ & 0.91$^{***}$ \\
\quad Diff. Costly Debt & -0.48$^{**}$ & -0.41$^{***}$ & -0.23 & -0.21 & -0.22 \\
\\[-1.4ex]
& \multicolumn{5}{c}{Degree of Shrinkage $r = 4$} \\ \cline{2-6} \\[-1.8ex]
IRF Spending: & & & & \\
\quad Baseline States & 1.06$^{***}$ & 2.22$^{***}$ & 2.68$^{***}$ & 2.41$^{***}$ & 0.21 \\
\quad Diff. Costly Debt & 0.16 & -0.58$^{**}$ & -2.64$^{***}$ & -2.65$^{***}$ & -0.33 \\
IRF Output:  & & & & \\
\quad Baseline States & 1.03$^{***}$ & 2.25$^{***}$ & 2.56$^{***}$ & 2.39$^{***}$ & 0.94 \\
\quad Diff. Costly Debt & -0.30 & -1.36$^{***}$ & -2.40$^{***}$ & -2.36$^{***}$ & -0.90 \\
Multipliers : & & & & \\
\quad Baseline States & 0.95$^{***}$ & 0.96$^{***}$ & 0.87$^{***}$ & 0.90$^{***}$ & 0.91$^{***}$ \\
\quad Diff. Costly Debt & -0.28 & -0.42$^{***}$ & -0.23 & -0.22 & -0.22 \\
\bottomrule
\end{tabular}

            \label{table appendix_degree_shrinkage.tex}
            \begin{flushleft}
                \begin{small}
                    \textit{Note:}  
                    This table shows the state-dependent impulse response functions (IRF) of output and public spending as well as the cumulative multipliers.
                    The IRF and multipliers are estimated with smooth local projections, for three different degree of shrinkage $r$.
                    Small Debt Cost are the point estimates IRF when the service of the debt is initially small. 
                    Diff. Costly Debt are the difference in point estimates IRF between states of large and small debt cost.  
                    The IRF estimates are measured in \% of GDP. The multipliers are measured in dollars. 
                    The forecast horizon ranges between 0 to 16 quarters.    
                    
                    $\star \  p < 0.10$,  $\star \star \  p < 0.05$,  $\star \star \star \ p < 0.01$       
                \end{small}
            \end{flushleft}
        \end{table}

        \FloatBarrier

\newpage
\FloatBarrier
\section{Alternative states definition}
\label{appendix section Alternative states definition}
    
    In this section, I show that the results of the paper are robust to using alternative definitions of the state variable. 

    \FloatBarrier
    \subsection{Smoothness of the transition}
    \label{appendix subsection smoothness of transition}

        The baseline state variable is defined by comparing the fiscal cost of public debt to its linear time trend, assuming a smooth transition between states of large and small debt service.
        This requires to take a stand on a numerical value for the speed of the transition $\gamma$.
        Table \ref{appendix table smoothness transition} shows that the main results of the paper is robust to alternative values on the speed of the transition, but also to using a dummy state variable, that is to assume instantaneous transition between both states. 

    \FloatBarrier
    \subsection{Increase state frequency}
    \label{appendix subsection Increase state frequency}

        Defining the state variable using a linear trend generates few observations of different states. 
        Intuitively, comparing an historical time series to its linear trend captures only slow-moving changes. 
        Hence, changes of regimes are rarely observed. 
        To generate more variations in the state variable, I compare the fiscal cost of public debt to its HP-filtered trend, using different values for the smoothness parameter of the filter $\lambda$. 
        The smaller this parameter, the more volatile the time trend, and therefore the more changes in the state variable. 
        As shown in table \ref{table appendix_hp_filter.tex}, adopting a more volatile state does not alter the main results of the paper.

        \begin{table}[h!]
            \centering
            \caption{Robustness: Smoothness of Transitions}
            \vspace{0.2cm} 
                \begin{tabular}{@{\extracolsep{10pt}}lccccc}
\toprule
\bottomrule
\\[-1.4ex]
 & $h = 0$ & $h = 4$ & $h = 8$ & $h=12$ & $h=16$ \\
\midrule
& \multicolumn{5}{c}{Smoothness $\gamma = 5$} \\ \cline{2-6} \\[-1.8ex]
IRF Spending: & & & & \\
\quad Small Debt Cost & 1.03$^{***}$ & 2.25$^{***}$ & 2.80$^{***}$ & 2.50$^{***}$ & 0.27 \\
\quad Diff. Costly Debt & 0.65$^{***}$ & -1.08$^{***}$ & -2.95$^{***}$ & -2.95$^{***}$ & -0.22 \\
IRF Output:  & & & & \\
\quad Small Debt Cost & 1.16$^{***}$ & 2.37$^{***}$ & 2.77$^{***}$ & 2.54$^{***}$ & 1.05$^{*}$ \\
\quad Diff. Costly Debt & -0.29 & -1.76$^{***}$ & -2.80$^{***}$ & -2.76$^{***}$ & -0.98 \\
Multipliers : & & & & \\
\quad Small Debt Cost & 1.07$^{***}$ & 1.00$^{***}$ & 0.90$^{***}$ & 0.93$^{***}$ & 0.96$^{***}$ \\
\quad Diff. Costly Debt & -0.44 & -0.49$^{***}$ & -0.25 & -0.25 & -0.27 \\
\\[-1.4ex]
& \multicolumn{5}{c}{Smoothness $\gamma = 15$} \\ \cline{2-6} \\[-1.8ex]
IRF Spending: & & & & \\
\quad Small Debt Cost & 1.10$^{***}$ & 2.22$^{***}$ & 2.69$^{***}$ & 2.32$^{***}$ & 0.21 \\
\quad Diff. Costly Debt & 0.46$^{***}$ & -0.62$^{**}$ & -2.39$^{***}$ & -2.57$^{***}$ & -0.21 \\
IRF Output:  & & & & \\
\quad Small Debt Cost & 1.13$^{***}$ & 2.21$^{***}$ & 2.56$^{***}$ & 2.30$^{***}$ & 0.90$^{*}$ \\
\quad Diff. Costly Debt & -0.27 & -1.32$^{***}$ & -2.26$^{***}$ & -2.24$^{***}$ & -0.76 \\
Multipliers : & & & & \\
\quad Small Debt Cost & 1.01$^{***}$ & 0.95$^{***}$ & 0.86$^{***}$ & 0.89$^{***}$ & 0.90$^{***}$ \\
\quad Diff. Costly Debt & -0.37 & -0.40$^{***}$ & -0.23 & -0.22 & -0.23 \\
\\[-1.4ex]
& \multicolumn{5}{c}{Dummy State Variable} \\ \cline{2-6} \\[-1.8ex]
IRF Spending: & & & & \\
\quad Small Debt Cost & 1.10$^{***}$ & 2.23$^{***}$ & 2.76$^{***}$ & 2.32$^{***}$ & 0.10 \\
\quad Diff. Costly Debt & 0.36$^{**}$ & -0.33 & -2.08$^{***}$ & -2.34$^{***}$ & -0.05 \\
IRF Output:  & & & & \\
\quad Small Debt Cost & 1.13$^{***}$ & 2.16$^{***}$ & 2.57$^{***}$ & 2.31$^{***}$ & 0.78 \\
\quad Diff. Costly Debt & -0.33 & -0.99$^{***}$ & -1.92$^{***}$ & -2.02$^{***}$ & -0.56 \\
Multipliers : & & & & \\
\quad Small Debt Cost & 1.01$^{***}$ & 0.93$^{***}$ & 0.84$^{***}$ & 0.87$^{***}$ & 0.87$^{***}$ \\
\quad Diff. Costly Debt & -0.38 & -0.36$^{***}$ & -0.19 & -0.17 & -0.17 \\
\bottomrule
\end{tabular}

            \label{appendix table smoothness transition}
            \begin{flushleft}
                \begin{small}
                    \textit{Note:}  
                    This table shows the state-dependent impulse response functions (IRF) of output and public spending as well as the cumulative multipliers.
                    The state variable is defined using different values for the speed of transition $\gamma$.    
                    Small Debt Cost are the point estimates IRF when the service of the debt is initially small. 
                    Diff. Costly Debt are the difference in point estimates IRF between states of large and small debt cost.  
                    The IRF estimates are measured in \% of GDP. The multipliers are measured in dollars. 
                    The forecast horizon ranges between 0 to 16 quarters.    
                    
                    $\star \  p < 0.10$,  $\star \star \  p < 0.05$,  $\star \star \star \ p < 0.01$       
                \end{small}
            \end{flushleft}
        \end{table}

        \FloatBarrier
        \begin{figure}[h!]
        \caption{The Fiscal Cost of Public Debt: HP-filtered Trend}
        \begin{center}
            \includegraphics{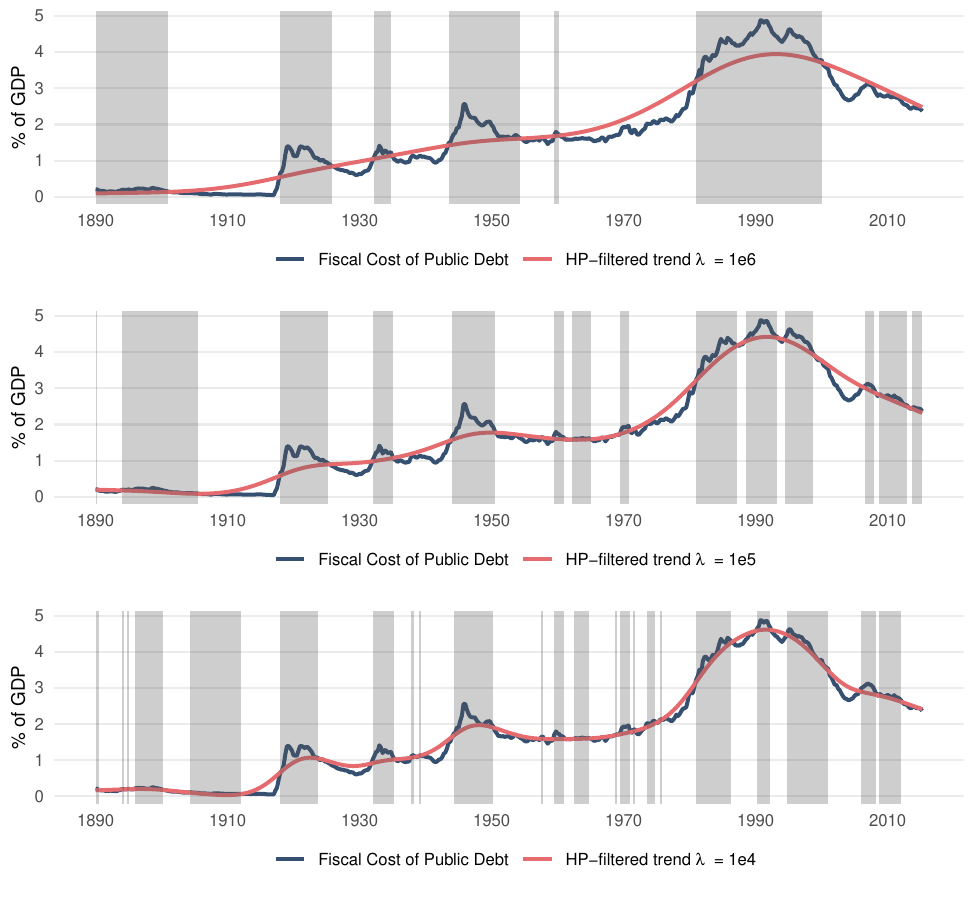} \label{fig ts_hp_states}
        \end{center}
        \vspace{-0.6cm}
        \begin{small}
            \textit{Note:} This figure shows the fiscal cost of public debt and its HP-filtered trend, for different smoothness parameter values. Shaded areas define periods where the series is larger than its trend. 
        \end{small}
        \end{figure}

        \begin{table}[h!]
            \centering
            \caption{Robustness: State Variable Volatility}
            \vspace{0.2cm} 
                \begin{tabular}{@{\extracolsep{10pt}}lccccc}
\toprule
\bottomrule
\\[-1.4ex]
 & $h = 0$ & $h = 4$ & $h = 8$ & $h=12$ & $h=16$ \\
\midrule
& \multicolumn{5}{c}{HP Filter $\lambda = 10^6$} \\ \cline{2-6} \\[-1.8ex]
IRF Spending: & & & & \\
\quad Baseline States & 1.06$^{***}$ & 2.24$^{***}$ & 2.86$^{***}$ & 2.50$^{***}$ & 0.28 \\
\quad Diff. Costly Debt & 0.45$^{***}$ & -0.34 & -2.25$^{***}$ & -2.59$^{***}$ & -0.26 \\
IRF Output:  & & & & \\
\quad Baseline States & 1.16$^{***}$ & 2.32$^{***}$ & 2.78$^{***}$ & 2.60$^{***}$ & 1.07$^{*}$ \\
\quad Diff. Costly Debt & -0.36 & -1.27$^{***}$ & -2.29$^{***}$ & -2.50$^{***}$ & -0.97 \\
Multipliers : & & & & \\
\quad Baseline States & 1.05$^{***}$ & 1.00$^{***}$ & 0.90$^{***}$ & 0.93$^{***}$ & 0.95$^{***}$ \\
\quad Diff. Costly Debt & -0.43 & -0.47$^{***}$ & -0.29$^{**}$ & -0.28$^{*}$ & -0.29 \\
\\[-1.4ex]
& \multicolumn{5}{c}{HP Filter $\lambda = 10^5$} \\ \cline{2-6} \\[-1.8ex]
IRF Spending: & & & & \\
\quad Baseline States & 1.07$^{***}$ & 2.28$^{***}$ & 2.89$^{***}$ & 2.59$^{***}$ & 0.43 \\
\quad Diff. Costly Debt & 0.50$^{***}$ & -0.92$^{***}$ & -2.58$^{***}$ & -2.60$^{***}$ & -0.32 \\
IRF Output:  & & & & \\
\quad Baseline States & 1.17$^{***}$ & 2.27$^{***}$ & 2.73$^{***}$ & 2.62$^{***}$ & 1.09$^{**}$ \\
\quad Diff. Costly Debt & -0.34 & -1.51$^{***}$ & -2.44$^{***}$ & -2.52$^{***}$ & -1.01$^{*}$ \\
Multipliers : & & & & \\
\quad Baseline States & 1.05$^{***}$ & 0.96$^{***}$ & 0.87$^{***}$ & 0.89$^{***}$ & 0.91$^{***}$ \\
\quad Diff. Costly Debt & -0.43 & -0.42$^{***}$ & -0.20 & -0.19 & -0.21 \\
\\[-1.4ex]
& \multicolumn{5}{c}{HP Filter $\lambda = 10^4$} \\ \cline{2-6} \\[-1.8ex]
IRF Spending: & & & & \\
\quad Baseline States & 1.05$^{***}$ & 2.16$^{***}$ & 2.59$^{***}$ & 2.84$^{***}$ & 0.51 \\
\quad Diff. Costly Debt & 0.22$^{**}$ & -1.03$^{***}$ & -2.53$^{***}$ & -2.62$^{***}$ & -0.01 \\
IRF Output:  & & & & \\
\quad Baseline States & 1.16$^{***}$ & 2.24$^{***}$ & 2.60$^{***}$ & 2.48$^{***}$ & 1.13$^{**}$ \\
\quad Diff. Costly Debt & -0.41 & -1.92$^{***}$ & -2.66$^{***}$ & -2.52$^{***}$ & -1.05$^{*}$ \\
Multipliers : & & & & \\
\quad Baseline States & 1.03$^{***}$ & 0.99$^{***}$ & 0.93$^{***}$ & 0.93$^{***}$ & 0.94$^{***}$ \\
\quad Diff. Costly Debt & -0.47 & -0.55$^{***}$ & -0.40$^{***}$ & -0.41$^{**}$ & -0.45 \\
\bottomrule
\end{tabular}

            \label{table appendix_hp_filter.tex}
            \begin{flushleft}
                \begin{small}
                    \textit{Note:}  
                    This table shows the state-dependent impulse response functions (IRF) of output and public spending as well as the cumulative multipliers.
                    The state variable is defined using a HP-filter with different value for its smoothness parameter $\lambda$.
                    Small Debt Cost are the point estimates IRF when the service of the debt is initially small. 
                    Diff. Costly Debt are the difference in point estimates IRF between states of large and small debt cost.  
                    The IRF estimates are measured in \% of GDP. The multipliers are measured in dollars. 
                    The forecast horizon ranges between 0 to 16 quarters.    
                    
                    $\star \  p < 0.10$,  $\star \star \  p < 0.05$,  $\star \star \star \ p < 0.01$       
                \end{small}
            \end{flushleft}
        \end{table}

\FloatBarrier
\section{Further robustness}
\label{Further robustness}

        \begin{table}[h!]
            \centering
            \caption{Robustness: Number of Lags}
            \vspace{0.2cm} 
                \begin{tabular}{@{\extracolsep{10pt}}lccccc}
\toprule
\bottomrule
\\[-1.4ex]
 & $h = 0$ & $h = 4$ & $h = 8$ & $h=12$ & $h=16$ \\
\midrule
& \multicolumn{5}{c}{Lags $p = 2$} \\ \cline{2-6} \\[-1.8ex]
IRF Spending: & & & & \\
\quad Baseline States & 1.08$^{***}$ & 2.28$^{***}$ & 2.86$^{***}$ & 2.53$^{***}$ & 0.37 \\
\quad Diff. Costly Debt & 0.57$^{***}$ & -0.84$^{***}$ & -2.78$^{***}$ & -2.83$^{***}$ & -0.41 \\
IRF Output:  & & & & \\
\quad Baseline States & 0.86$^{***}$ & 1.96$^{***}$ & 2.38$^{***}$ & 2.17$^{***}$ & 0.79 \\
\quad Diff. Costly Debt & 0.07 & -1.16$^{***}$ & -2.14$^{***}$ & -2.15$^{***}$ & -0.68 \\
Multipliers : & & & & \\
\quad Baseline States & 0.79$^{***}$ & 0.77$^{***}$ & 0.74$^{***}$ & 0.77$^{***}$ & 0.80$^{***}$ \\
\quad Diff. Costly Debt & -0.09 & -0.24$^{*}$ & -0.11 & -0.12 & -0.14 \\
\\[-1.4ex]
& \multicolumn{5}{c}{Lags $p = 8$} \\ \cline{2-6} \\[-1.8ex]
IRF Spending: & & & & \\
\quad Baseline States & 1.03$^{***}$ & 2.18$^{***}$ & 2.55$^{***}$ & 2.67$^{***}$ & 0.24 \\
\quad Diff. Costly Debt & 0.08 & -0.59$^{*}$ & -2.27$^{***}$ & -2.55$^{***}$ & 0.03 \\
IRF Output:  & & & & \\
\quad Baseline States & 1.26$^{***}$ & 2.50$^{***}$ & 2.88$^{***}$ & 2.60$^{***}$ & 1.13$^{**}$ \\
\quad Diff. Costly Debt & -0.26 & -1.42$^{***}$ & -2.32$^{***}$ & -2.46$^{***}$ & -1.01$^{*}$ \\
Multipliers : & & & & \\
\quad Baseline States & 1.16$^{***}$ & 1.08$^{***}$ & 0.95$^{***}$ & 0.95$^{***}$ & 0.96$^{***}$ \\
\quad Diff. Costly Debt & -0.35 & -0.46$^{***}$ & -0.21 & -0.19 & -0.26 \\
\bottomrule
\end{tabular}

            \label{table appendix_nlags.tex}
            \begin{flushleft}
                \begin{small}
                    \textit{Note:}  
                    This table shows the state-dependent impulse response functions (IRF) of output and public spending as well as the cumulative multipliers, for different number of lags for control variables.   
                    Small Debt Cost are the point estimates IRF when the service of the debt is initially small. 
                    Diff. Costly Debt are the difference in point estimates IRF between states of large and small debt cost.  
                    The IRF estimates are measured in \% of GDP. The multipliers are measured in dollars. 
                    The forecast horizon ranges between 0 to 16 quarters.    
                    
                    $\star \  p < 0.10$,  $\star \star \  p < 0.05$,  $\star \star \star \ p < 0.01$       
                \end{small}
            \end{flushleft}
        \end{table}

\newpage
\FloatBarrier
\section{Instrument Relevance for Multipliers}
\label{appendix section Instrument Relevance for Multipliers}

\begin{figure}[h!]
      \caption{Instruments Relevance}
      \begin{center}
        \includegraphics{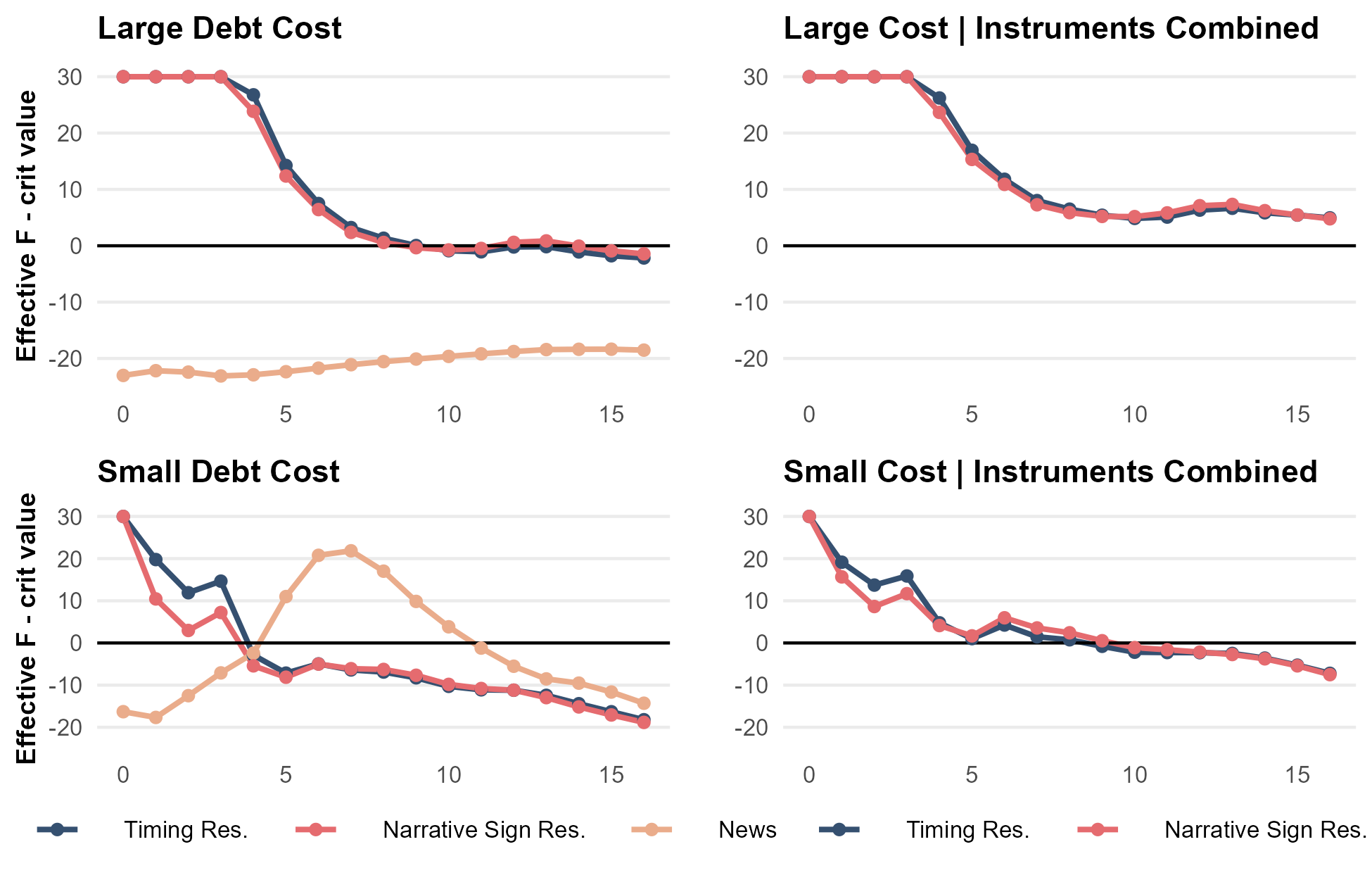} \label{fig appendix f_stat_debt_cost}
      \end{center}
      \vspace{-0.6cm}
      \begin{small}
        \textit{Note:}  
        This figure shows the difference between the effective F statistics and critical values of \cite{oleaRobustTestWeak2013} to test the first stage of the multipliers regression, with different government spending shocks. 
        The test is conducted in two possible states of the economy: small and large debt cost.The left panel shows the result of the test with the three different shocks taken separately in the first stage. The right panel shows the result of the test when either the timing or the narrative sign restriction shocks are combined with \cite{rameyIdentifyingGovernmentSpending2011} news shocks.
      \end{small}
    \end{figure}

    The LP-IV model \eqref{eq state dependent multiplier} allows to assess the relevance of the spending shock time series as instruments to public spending. 
    As standard and smooth local projections yield auto-correlated errors, I use \cite{oleaRobustTestWeak2013} effective F-statistics and thresholds to test the first stage of the multipliers regression. 
    Figure \ref{fig appendix f_stat_debt_cost} shows the difference between the effective F statistics and the critical values of the first stage regression, for different spending shocks, in the two states of small and large debt cost. 
    A value greater than 0 means that the shock is a relevant instrument for public spending in the given state of the economy.
    The left panel of figure \ref{fig appendix f_stat_debt_cost} reveals that shocks identified through SVAR (either with timing or narrative sign restrictions) are relevant at all horizon when the cost of debt is initially large.
    However, during episodes of small debt service, the relevance of SVAR shocks declines at long horizon.
    On the other hand, \cite{rameyIdentifyingGovernmentSpending2011} news shocks are not relevant when the cost of debt is large, and are relevant only at medium horizon when the cost of debt is small.
    However, combining news shocks with either timing or narrative sign restriction shocks improve the long term relevance of the first stage, as shown in the right panel of figure \ref{fig appendix f_stat_debt_cost}.

\end{document}